\documentclass[letterpaper,twocolumn,10pt]{article}
\usepackage{hyperref}
\usepackage{usenix2019_v3}

\usepackage{booktabs} 
\usepackage{todonotes}
\usepackage{fancyhdr}
\usepackage{enumitem}
\usepackage{amsmath}

\usepackage{epstopdf}
\usepackage{xcolor,colortbl}
\usepackage{color}
\usepackage{algorithm}
\usepackage[noend]{algpseudocode}
\usepackage{graphicx}
\graphicspath{{figures/}}
\usepackage{subfig}
\usepackage{cite}
\usepackage{tikz}
\usetikzlibrary{matrix}
\usepackage{filecontents}
\usepackage{xstring,etoolbox,catchfile}

\usepackage{authblk}

\usepackage{pgfplots} 
\usepackage{environ}
\usepackage{scalefnt}
\pgfplotsset{compat=newest} 
\pgfplotsset{plot coordinates/math parser=false} 
\newlength\figureheight 
\newlength\figurewidth 

\begin{document}
\date{}

\title{RELOAD+REFRESH: Abusing Cache Replacement Policies to Perform Stealthy Cache Attacks}

\author[1]{Samira Briongos}
\author[1]{Pedro Malag\'on}
\author[1]{Jos\'e M. Moya}
\author[2,3]{Thomas Eisenbarth}
\affil[1]{Integrated Systems Laboratory, Universidad Polit\'ecnica de Madrid, Madrid, Spain}
\affil[2]{University of L\"ubeck, L\"ubeck, Germany}
\affil[3]{Worcester Polytechnic Institute, Worcester, MA, USA}
\renewcommand\Authands{ and }

\maketitle

\begin{abstract}
Caches have become the prime method for unintended information extraction across logical isolation boundaries. Even Spectre and Meltdown rely on the cache side channel, as it provides great resolution and is widely available on all major CPU platforms. As a consequence, several methods to stop cache attacks by detecting them have been proposed. Detection is strongly aided by the fact that observing cache activity of co-resident processes is not possible without altering the cache state and thereby forcing evictions on the observed processes.
In this work we show that this widely held assumption is incorrect. Through clever usage of the cache replacement policy it is possible to track a victims process cache accesses \emph{without forcing evictions} on the victim's data. Hence, online detection mechanisms that rely on these evictions can be circumvented as they do not detect be the introduced RELOAD+REFRESH attack. The attack requires a profound understanding of the cache replacement policy. We present a methodology to recover the replacement policy and apply it to the last five generations of Intel processors. We further show empirically that the performance of RELOAD+REFRESH on cryptographic implementations is comparable to that of other widely used cache attacks, while its detectability becomes extremely difficult, due to the negligible effect on the victims cache access pattern. 

\end{abstract}

\section{Introduction}

The microarchitecture of modern CPUs shares resources among concurrent processes. This sharing may result in unintended information flows between concurrent processes. Microarchitectural attacks, which exploit these information flows, have received a lot of attention in academia, industry and, with Spectre and Meltdown~\cite{Kocher2018spectre,lipp2018meltdown}, even in the public news. The OS or the hypervisor in virtual environments provide strict logical isolation among processes to enable secure multithreading. Yet, a malicious process can intentionally create contention to gain information about co-resident processes. Exploitable hardware resources include the branch prediction unit \cite{Aciicmez2007,AciicmezBranch,Aciicmez2}, the DRAM~\cite{rowhammer1,6853210,197211} and the cache \cite{GullaschEtAl2011,Oren2015,YaromEtAl2014,Irazoqui14,OsvikEtAl2006,lastlevel}. Last level caches (LLC) provide very high temporal and spatial resolution to observe and track memory access patterns. As a consequence, any code that generates cache utilization patterns dependent on secret data is vulnerable.
Cache attacks can trespass VM boundaries to infer secret keys from neighboring processes or VMs~\cite{Ristenpart09,Inci2016}, break security protocols~\cite{IrazoquiEtAl2015,ronen19} or compromise the end users privacy~\cite{Oren2015}, but they can leak information from within a victim memory address space~\cite{Kocher2018spectre} when combined with other techniques. 

Different techniques have been proposed for detection and/or mitigation of cache and other microarchitectural attacks due to the great threat they pose. On the one hand, hardware countermeasures take years to integrate and deploy, may induce performance penalties and currently, are not implemented. On the other hand, other proposals that are meant for cloud hypervisors~\cite{180212,li2014stopwatch,Varadarajan14} and only require making small modifications to the kernel configuration are neither implemented presumably due to the overhead they entail. 

The only solution that seems practical for users that want to protect themselves against these kind of attacks, is to detect ongoing attacks and then react in some way. To this end, different proposals~\cite{ChiappettaEtAl2016,Payer2016,Zhang2016,Briongos_2018,Kulah2018} use hardware performance counters (HPCs), which are special registers available in all modern CPUs, that monitor hardware events such as cache misses. The most recent proposals are able to detect even attacks specially designed to bypass previous countermeasures~\cite{GrussEtAl2016}. The common assumption in these works is that the \emph{attacker induces measurable effects} on the victim. We, on the contrary, demonstrate that it is possible to obtain information from the victim while keeping its data in the cache and, consequently, not significantly altering its behavior, thus making attack detection difficult.  

\paragraph{Our Contribution:} 
We analyze the replacement policy of current Intel CPUs and identify a new strategy which allows an attacker to monitor cache set accesses without forcing evictions of the victim's data, thereby creating a new and \emph{stealthy} cache-based microarchitectural attack. 
To achieve this goal, we perform the first full reverse engineering of different replacement policies present in various generations of Intel Core processors. We propose a technique that can be extended to study replacement policies of other processors. Using this technique, we demonstrate that it is possible to accurately predict which element of the set will be replaced in case of a cache miss. Then, we show that it is possible to exploit these deterministic cache replacement policies to derive a sophisticated cache attack: RELOAD+REFRESH, which is able to monitor the memory accesses of the desired victim without generating cache misses. As a proof of concept, we demonstrate the feasibility and quantify the performance of RELOAD+REFRESH by retrieving the key of a T-Table implementation of AES and attacking the square and multiply version of RSA. We prove that our approach is at least as accurate as other state-of-the-art cache attacks. Even when adding extra steps to recover the information, the attack has still enough resolution (similar to PRIME+PROBE) to trace the victim memory accesses with high accuracy. Thus, our work reveals the need for new detection mechanisms or countermeasures. To sum up, this work:

\begin{itemize}
    \item introduces a methodology to test different replacement policies in modern caches.
    \item uncovers the replacement policy currently implemented in modern Intel Core processor generations, from fourth to eight generation.
    \item expands the understanding of modern caches and lays the basis for improving traditional cache attacks.
    \item presents RELOAD+REFRESH, a new attack that exploits Intel cache replacement policies to extract information referring to a victim memory accesses.
    \item shows that the proposed attack causes negligible cache misses on the victim, which renders it undetectable by state of the art countermeasures. 
\end{itemize}

\section{Background and related work}\label{sec:background}
\subsection{Cache architecture}
CPU caches are small banks of fast memory located between the CPU cores and the RAM. As they are placed on the CPU die and close to the cores, they have low access latencies and thus reduce memory access times observed by the processor, improving the overall performance. Modern processors include cache memories that are hierarchically organized; low level caches (L1 and L2) are core private, smaller and closer to the processor, whereas the last level cache (LLC or L3) is bigger and shared among all the cores.

Intel processors traditionally have included L3 inclusive caches: all the data which is present in the private lower caches has to be in the shared L3 cache. This approach makes cache coherence much easier to implement. However, due to cache attacks, the newest Intel Skylake Server micro architecture considers using a non-inclusive Last Level Cache~\cite{intelman}.

In most modern processors caches are W-way set-associative. The cache is organized into multiple sets ($S$), each of them containing $W$ lines of usually 64 bytes of data. The set in which each line is placed is derived from its address. The address bits are divided into offset (lowest-order bits used to locate data within a line), index (\textit{log$_2$(S)} consecutive bits starting from the offset bits that address the set) and tag (remaining bits which identify if the data is cached). 

\subsection{Cache replacement policies}

When the processor requests some data, it first tries to retrieve this data from the cache (it starts looking in the lowest levels up to the last level). In the event of a \textit{cache hit} the data is loaded from the cache. On the contrary, in the event of a \textit{cache miss}, the data is retrieved from the main memory and it is also placed in the cache assuming that it will be re-used in the near future. If there is no free room in the cache set, the memory controller has to decide which element in the cache has to be evicted. Since the processor is stalled for several cycles whenever there is a \textit{cache miss}, the decision of which data is evicted and which data stays is crucial for the performance.

Many replacement policies are possible including, for example, FIFO (first in first out), LRU (least recently used) or its approximations such as NRU \cite{nru_pol} (not recently used), LFU (least frequently used), CLOCK \cite{clockPol}(keeps a circular list of the elements) or even pseudo-random replacement policies. Modern high performance processors implement approximations to LRU, because LRU is hard to implement (it requires complex hardware to track each access). 

LRU or pseudo LRU policies have demonstrated to perform well in most situations. Nevertheless, LRU policy behaves poorly for memory-intensive workloads whose working set is bigger than the available cache size or for scans (bursts of one-time access requests). As a result, adaptive algorithms, which are capable to adapt themselves to changes in the workloads, came up. In 2003, Megiddo el al. \cite{Megiddo2003} proposed ARC (Adaptive Replacement Cache) a hybrid of LRU and LFU. One year later, Bansal et al. \cite{Bansal2004} presented their solution based on LFU and CLOCK, which they named CAR (Clock with Adaptive Replacement).

In 2007 Quereshi et al. \cite{Qureshi07} suggested that performance could be improved by changing the insertion policy while maintaining the eviction policy. LIP (LRU Insertion Policy) consists in inserting each new piece of data in the LRU position whereas BIP (Bimodal Insertion Policy) most of the times places the new data in the MRU position and sometimes inserts it in the LRU position. In order to decide which of the two policies behaves better, they proposed a dynamic insertion policy (DIP). DIP chooses between LIP and BIP depending on which one incurs fewer misses.

In 2010, Jaleel et al. \cite{Jaleel10} proposed a cache replacement algorithm that makes use of Re-reference Interval Prediction (RRIP). By using 2 bits per cache line, RRIP predicts if a cache line is going to be re-referenced in the near future. In case of eviction, the line with the longest interval prediction will be selected. Analogously to Quereshi et al, they presented two different approaches: Static RRIP (SRRIP) which inserts each new block with an intermediate re-reference, and Bimodal RRIP (BRRIP) which inserts most blocks with a distant re-reference interval and sometimes with an intermediate re-reference interval. They also proposed using set dueling to decide which policy fits better for the running application (Dynamic RRIP or DRRIP).

Regarding to Intel processors, their replacement policy is undocumented and consequently unknown. All that is officially known is the name of the policy: "Quad-Age LRU" \cite{7476478}. The first serious attempt to reveal the cache replacement policy of different processors was made by Abel et al. \cite{reversePol1}. In their work, they were able to uncover the replacement policy of an Intel Atom D525 processor and to infer a pseudo-LRU policy in an Intel Core 2 Duo E6300 processor. However, they could not determine the eviction policy in the other machines (Intel Core 2 Duo E6750 and E8400) they used for the experiments. Later on, Henry \cite{pagewalk-coherence} showed that Intel processors seem to implement a dynamic insertion or eviction policy, but he did not provide further details about the replacement policy. 

Gruss et al.\cite{Gruss2016Row} studied cache eviction strategies on recent Intel CPUs in order to replace the \texttt{clflush} instruction and build a remote Rowhammer attack. As they mention, their work is not strictly a reverse engineering of the replacement policy, rather they test access patterns to find the best eviction strategy. To the best of our knowledge, our work is the first one that provides a comprehensive description of the replacement policies implemented on modern Intel processors.

\subsection{Cache attacks}

Cache attacks monitor the utilization of the cache (the sequence of cache hits and misses) to retrieve information about a co-resident victim. Whenever the pattern of memory accesses of a security-critical piece of software depends on the actual value of sensible data, such as a secret key, this sensitive data can be deduced by an attacker and will no longer be private.
Traditionally cache attacks have been grouped into three categories~\cite{Ge2018}: FLUSH+RELOAD, PRIME+PROBE and EVICT+TIME. From those, the  FLUSH+RELOAD and the PRIME+PROBE attacks (and their variants) out-stand over the rest due to their higher resolution.

Both attacks target the LLC, selecting one memory location that is expected to be accessed by the victim process. They consist of three stages: \texttt{initialization} (the attacker prepares the cache somehow), \texttt{waiting} (the attacker waits while the victim executes) and \texttt{recovering} (the attacker checks the state of the cache to retrieve information about the victim).

\subsubsection{FLUSH+RELOAD}

This attack relies on the existence of shared memory. Thus, it requires memory deduplication to be enabled. Deduplication is an optimization technique designed to improve memory utilization by merging duplicate memory pages. Using the \texttt{clflush} instruction the attacker removes the target lines from the cache, then waits for the victim process to execute (or an equivalent estimated time) and finally measures the time it takes to reload the previously flushed data. Low reload times mean the victim has used the data.

It was first introduced in~\cite{GullaschEtAl2011}, and was later extended to target the LLC to retrieve cryptographic keys, TLS protocol session messages or keyboard keystrokes across VMs~\cite{YaromEtAl2014,GrussEtAl2015,IrazoquiEtAl2015}. Further, Zhang et al.~\cite{ZhangEtAlPaaS} showed that it was applicable in several commercial PaaS clouds. 

Relying on the \texttt{clflush} instruction and with the same requirements as FLUSH+RELOAD, Gruss et al. ~\cite{GrussEtAl2016} proposed the FLUSH+FLUSH attack. It was intended to be stealthy and bypass existing monitoring systems. This variant recovers the information by measuring the execution time of the \texttt{clflush} instruction instead of the reload time, thus avoiding direct cache accesses and, as a consequence, detection. However, recent works have demonstrated that it is detectable \cite{Briongos_2018,Kulah2018}.

\subsubsection{PRIME+PROBE}

Contrary to the FLUSH+RELOAD attack, PRIME+PROBE is agnostic to special OS features in the system. Therefore, it can be applied in virtually every system. Moreover, it can recover information from dynamically allocated data. To do so, the attacker first fills or primes the cache set in which the victim data will be placed (initialization stage). Then, he waits and finally probes the desired set looking for time variations that carry information about the victim activity. 

This attack was first proposed for the L1 data cache in~\cite{OsvikEtAl2006} and later was expanded to the L1 instruction cache~\cite{aciiccmez2008}. These approaches required both victim and attacker to share the same core, which diminishes practicality. However, it has been recently shown to be applicable to LLC. Researchers have bypassed several difficulties to target the LLC, as retrieving its complex address mapping~\cite{Maurice_2015,Yarom2015MappingTI,7302337}, and recovered cryptographic keys or keyboard typed keystrokes~\cite{lastlevel,sca,DBLP:conf/uss/LippGSMM16}. Even further, the PRIME+PROBE attack was used to retrieve a RSA key in the Amazon EC2 cloud~\cite{Inci2016}.

In case a defense system tries to either restrict access to the timers \cite{Martin2912,197223} or to generate noise that could hide timing information, cache attacks are less likely to succeed. The PRIME+ABORT attack ~\cite{disselkoen17} overcomes this difficulty. It exploits Intel's implementation of Hardware Transactional Memory (TSX) to retrieve the information about cache accesses. It first starts a transaction to prime the targeted set, waits and finally it may or may not receive and abort depending on whether the victim has or has not accessed this set. 

\subsection{Countermeasures}

Researchers have tackled the problem of mitigating cache attacks from different perspectives.  
Several proposals suggest limiting the access to the shared resources that can be exploited to infer information about a victim by modifying the underlying hardware \cite{Liu2014,Wang2007}. While effective, these hardware countermeasures take years to deploy and it is not likely that any CPU manufacturer will deploy them. 

System-level software approaches, on the other hand, require modification of the current cloud infrastructure or the Linux kernel. STEALTHMEM ~\cite{180212} uses private virtual pages that ensure the data located in them is not evicted from the cache and avoid mapping any other page with these private virtual pages. CATalyst~\cite{7446082_cata} uses Intel Cache Allocation Technology (CAT), which is a technology that enables system administrators to control how cores allocate data into the LLC, to mitigate cache attacks. CACHEBAR \cite{Zhou2016} designs a memory management subsystem that dynamically changes the amount or lines per cache set that a security domain can occupy to defeat PRIME+PROBE attacks and changes the state of the pages to avoid FLUSH+RELOAD. These countermeasures are more plausible than the previous ones, however no cloud provider or OS is implementing them, probably because of the performance penalties they incur. 

For these reasons, we believe that the only countermeasures that an attacker may have to face when trying to retrieve information from a victim, are detection based countermeasures which can be implemented at user level. Cache attacks exploit the side effects of running a program in certain hardware to gain information from it, and similarly, these countermeasures employ monitoring mechanisms to detect such attacks. Detection systems can use time measurements \cite{Briongos2016}, hardware performance counters \cite{ChiappettaEtAl2016,Briongos_2018,Zhang2016,Kulah2018} or place data in transactional regions \cite{203672} defined with the Intel TSX instructions. All of them measure the effect of the last level cache misses on the victim or on both the victim and the attacker. As a consequence, an attack that does not generate cache misses on the victim side would be undetectable by these systems.

A different approach to protect sensitive applications is to specifically design them to be secure against side-channels (no memory accesses depend on private information). Then, developers can use specific tools to ensure the binary of such applications does not leak information, even if it is under attack~\cite{ZanklEtAl2017,Wichelmann2018}. 
There are other tools, such as MASCAT \cite{Irazoqui:2018:MPM:3176258.3176316}, which use code analysis techniques to detect potential attacks before running a program, as most anti-viruses do. This kind of tools is effective before
malware distribution or execution. Their effectiveness is reduced in cloud environments where the attacker does not need to infect the victim. 

\section{Retrieval of Intel cache eviction policies}\label{retri}

This work focuses on the LLC. Since it is shared across cores, the attacks targeting the LLC are not limited to the situation in which the victim and the attacker share the same core. It is also possible to extract fine-grained information from it and many researchers are concerned about the attacks targeting the LLC. Attacks that assume a pseudo LRU eviction policy such as PRIME+PROBE or EVICT+RELOAD can benefit from the detailed knowledge of the eviction policy, as can benefit one attacker wishing to carry out a stealthy attack that does not cause cache misses on the victim. 

In order to be able to study the eviction policy, we have to ensure we can fill one set of the cache with our own data, access that data and force a miss when desired to observe which element of the set is evicted. For this reason, we have constructed an eviction set (a group of $w$ different addresses that map to one specific set in $w$-way set-associative caches) and what we call a conflicting set (a second eviction set that maps to exactly the same set and is composed of disjoint addresses). Previous works have shown how to statically recover the complex addressing function ~\cite{Maurice_2015,Yarom2015MappingTI,7302337}. However, we have decided to create both the eviction and conflicting sets dynamically \cite{lastlevel}. This approach is faster and general for all the processors involved in this work.

\renewcommand{\algorithmicrequire}{\textbf{Input:}}
\renewcommand{\algorithmicensure}{\textbf{Output:}}
\begin{algorithm}
\caption{Obtaining the conflicting set}\label{conf_ob}
\begin{algorithmic}[0]
\Require \textbf{Eviction\_set, Conflicting set candidates}
\Ensure \textit{\textbf{Conflicting\_set}}
\Function{getConflictingSet}{eviction\_set, candidates}
\State $conflicting\_set\gets \{\};$
\ForAll{$e \in$ candidates}
\State \emph{read\_all}(eviction\_set);
\State \emph{flush\_all}(eviction\_set);
\State \emph{read}($e$);
\State \emph{read\_all}(eviction\_set);
\State measure $time$ to read $e$;
\If{$time>threshold$}
\State $conflicting\_set\gets e;$
\If{$sizeof(conflicting\_set)==w$}
\State break;
\EndIf
\EndIf
\EndFor
\State \textbf{return} $conflicting\_set$
\EndFunction
\end{algorithmic}
\end{algorithm}

The eviction set was constructed following the procedure proposed by Liu et al. in \cite{lastlevel} (Algorithm 1). The procedure for obtaining the conflicting set follows the same principles and uses the eviction set that has just been constructed. This procedure is summarized in algorithm \ref{conf_ob}. First, we remove all the data from one set by accessing the eviction set and then, once the data is in the cache set, \textit{flushing} the whole eviction set. When the set is completely empty, we select one address among all the possible candidates (memory lines whose set index bits are equal to the eviction set index bits) for the conflicting set. 
Finally, we access the candidate address, the whole eviction set, and re-access the candidate again. If the reload time is higher than a threshold, we can conclude that the candidate address maps to the same set as the eviction set. Thus, it can be included in our conflicting set. We repeat the same procedure until the conflicting set has $w$ lines. 

For all the experiments, we have enabled the use of hugepages in our systems. Note that the order of the accesses is important to deduce the eviction policy. We enforce this order using \emph{mfence} instructions, which act as barriers that ensure all preceding load and store instructions have finished before any load or store instruction that follows \emph{mfence}.

\begin{table*}[th]
\caption{Details of the machines used in this work to retrieve their Replacement Policies}
\label{tab:gen}
\centering
\begin{tabular}{|c|c|c|c|c|c|}
\cline{1-6}
\textbf{Processor} & i7-4790 & i3-5010U & i7-6700K & i5-7600K & i7-8650U\\
\cline{1-6}
\textbf{Cores} & 4 & 2 & 4 & 4 & 4\\
\cline{1-6}
\textbf{Associativity} & 16 & 12 & 16 & 12 & 16\\
\cline{1-6}
\textbf{OS} & Centos 7.0 & Ubuntu 14 & Ubuntu 16 & Centos 7.0 & Debian 9.5\\
\cline{1-6}
\end{tabular}
\end{table*}

\subsection{Design of the experiments}

We have performed experiments in five different machines, each of them including an Intel processor from a different generation. Table \ref{tab:gen} presents a summary of the machines employed in this work. It includes the processor name, its number of cores, the associativity of the cache and the OS running on each machine. We have started by studying the processor of the fourth generation, which has been a common victim of published PRIME+PROBE attacks. We have finally covered from fourth to eighth generations.

Before conducting the experiments to disclose the eviction policy implemented in each of the used machines, we have performed some experiments intended to verify that no cached data is evicted in the event of a cache miss if there is free room in the set. The procedure is quite straightforward: for each of the sets, we first completely fill it with the data on its corresponding eviction set. Next, we randomly \emph{flush} one of these lines to ensure there is free room in the set, and we access one of the lines in the conflicting set to ensure there is going to be a cache miss. Finally, we check, by measuring times when re-accessing them, that all the lines in the eviction set (except for the one evicted) still reside in the cache. As expected, in all cases the incoming data was loaded in replacement of the \textit{flushed} line. 

Note that in a machine fully controlled by us, we can compare the actual evolution of the data in each of the sets with its theoretical evolution defined by an eviction policy during the runtime. This is the main idea of the procedure we propose to retrieve the replacement policy. Algorithm \ref{test_pol} summarizes this procedure. Each of the policies that has been tested had to be manually defined. We have evaluated true LRU, Tree PLRU, CLOCK, NRU, Static and Bimodal RRIP, self-defined policies using four control bits, etc. among many other possible cache eviction policies. After multiple experiments, we can conclude that the implemented policy is the defined policy which best matches the experimental observations.

Algorithm \ref{test_pol} tries to emulate by software the behavior of the hardware (of the cache). For this purpose, it uses two arrays of size $w$. On the one hand, \emph{address\_array} mimics the studied set, storing the memory addresses whose data is in the cache set. On the other hand, \emph{control\_array} contains the control bits used for deciding which address will be evicted in case of conflict. Additionally, we need to manually define one function that updates the content of the \emph{address\_array}, one function that updates the \emph{control\_array} and another one that provides the eviction candidate i.e. it returns the address of the element that will be evicted in case of conflict. These functions are defined based on the replacement policy.

As an example, we assume we want to test the NRU policy \cite{nru_pol}, which turns out to match the policy implemented in an Intel Xeon E5620 according to our experiments. According to its specification, NRU uses one bit per cache line, this bit is set whenever a cache line is accessed. If setting one bit implies that all the bits of a cache set will be equal to one, then all the bits (except for the one that has just being accessed) will be cleared. In case of conflict, NRU will remove from the cache one element whose control bit is equal to zero. Thus in our procedure, the control bits would be -1 (line empty), 0 (line not recently used), and 1 (line recently used). When a memory line is accessed, the \textit{update} function first checks if its address is already included in the \emph{address\_array}. If it is not, our function will add it to the \emph{address\_array} and set the corresponding bit in the \emph{control\_array}. On the contrary, the function only updates the \emph{control\_array}. The \textit{getEvictionCandidate} function will return one array position whose control bit value is -1, or, if no control bit is equal to -1, one whose control bit is equal to 0. In case multiple addresses have control bits equal to -1 or to 0, the function will return the first address whose control bits are -1 or 0, that it encounters when traversing the \emph{control\_array} from the beginning. Finally, after forcing a cache miss, the \textit{testDataEvicted()} returns the element truly evicted that we then compare with the predicted by the NRU policy (the output of \textit{getEvictionCandidate}).

\begin{figure}[hbt]
\centering
%
%
\definecolor{mycolor1}{rgb}{0.00000,0.44700,0.74100}%
\begin{tikzpicture}

\begin{axis}[%
width=2.7in,
height=1.0in,
at={(0.094in,0.424in)},
scale only axis,
xmin=0,
xmax=400,
xlabel style={font=\color{white!15!black}},
xlabel={Access times},
ymin=0,
ymax=2000000,
ylabel style={font=\color{white!15!black}},
ylabel={Number of samples},
axis background/.style={fill=white}
]
\addplot [color=mycolor1, line width=2.0pt, forget plot]
  table[row sep=crcr]{%
0	0\\
5	0\\
10	0\\
15	0\\
20	0\\
25	0\\
30	0\\
35	0\\
40	0\\
45	0\\
50	0\\
55	0\\
60	0\\
65	0\\
70	22276\\
75	1015347\\
80	614921\\
85	561895\\
90	187272\\
95	52330\\
100	303413\\
105	1830206\\
110	1376063\\
115	1192170\\
120	361984\\
125	44743\\
130	425\\
135	31\\
140	10\\
145	3\\
150	1\\
155	2\\
160	3\\
165	0\\
170	0\\
175	0\\
180	2\\
185	0\\
190	0\\
195	0\\
200	0\\
205	0\\
210	0\\
215	0\\
220	0\\
225	0\\
230	0\\
235	1\\
240	0\\
245	1\\
250	30\\
255	271\\
260	319\\
265	246\\
270	156\\
275	26\\
280	28\\
285	4\\
290	0\\
295	0\\
300	0\\
305	0\\
310	1\\
315	10\\
320	51\\
325	29\\
330	96\\
335	1148\\
340	119752\\
345	141602\\
350	182328\\
355	74255\\
360	41972\\
365	4602\\
370	600\\
375	270\\
380	262\\
385	234\\
390	253\\
395	211\\
400	248\\
405	243\\
410	238\\
415	246\\
420	247\\
425	232\\
430	252\\
435	240\\
440	222\\
445	269\\
450	221\\
455	257\\
460	215\\
465	274\\
470	194\\
475	268\\
480	279\\
485	1082\\
490	1379\\
495	1660\\
500	51879\\
};
\end{axis}

\begin{axis}[%
width=2.7in,
height=1.4in,
at={(0.094in,0.424in)},
scale only axis,
xmin=0,
xmax=1,
ymin=0,
ymax=1,
axis line style={draw=none},
ticks=none,
axis x line*=bottom,
axis y line*=left
]
\node[below, align=center]
at (rel axis cs:0.50,0.74) {L3 cache accesses};
\node[below, align=center]
at (rel axis cs:0.126,0.70) {Low level\\accesses};
\node[below, align=center, draw=white]
at (rel axis cs:0.83,0.40) {Main memory\\accesses};
\end{axis}
\end{tikzpicture}%
\vspace{-20 pt}
\caption{Distribution of the access times to different data. These times depend on which memory it was located.} \label{relo_times}
\end{figure}
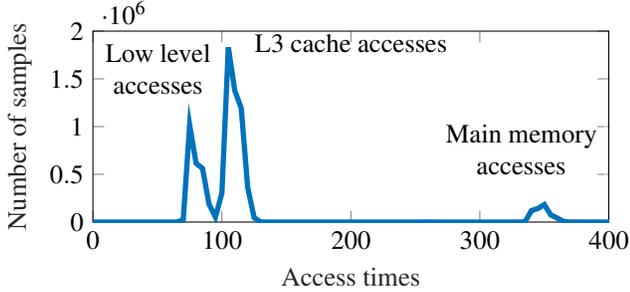

We have noticed that only accesses to the LLC update the values of the control bits of the accessed element. That is, if the data is located in L1 or L2 caches when requested (reload time lower than ll\_threshold), we do not update the values in the \emph{control\_array}. Figure~\ref{relo_times} shows the distinction between accesses to low and last level caches based on reload times.

\renewcommand{\algorithmicrequire}{\textbf{Input:}}
\renewcommand{\algorithmicensure}{\textbf{Output:}}
\begin{algorithm}[th]
\caption{Test of
the desired eviction policy}\label{test_pol}
\begin{algorithmic}[0]
\Require \textbf{Eviction\_set, Conflicting\_set}
\Ensure \textbf{Accuracy of the policy} \Comment{hits/trials}
\Function{testPolicy}{eviction\_set, conflicting\_set}
\State $hits=0;$
\While{$i \leq num\_experiments$}
\State $j=0,$i++;
\State $\emph{control\_array}\gets \{\};  \emph{address\_array}\gets \{\};$
\State \emph{initialize\_set()}; \Comment{Fills address and control arrays}
\State $lim=random();$
\While{$j \leq lim $}
\State \emph{mfence}; j++;
\State $next\_data=eviction\_set[random()];$ 
\State measure $time$ to read $next\_data$;
\If{$time \geq ll\_threshold$} \Comment{LLC access}
\State $update(\emph{control\_array},next\_data);$
\EndIf
\EndWhile
\State $conf\_element=conflicting\_set[random()];$
\State \textbf{read}($conf\_element$); \Comment{Force miss}
\State candidate=$getEvictionCandidate();$
\If{($\mathit{testDataEvicted()==}$candidate)} 
\State $hits$++;
\EndIf
\EndWhile
\State \textbf{return} $hits \slash num\_experiments;$
\EndFunction
\end{algorithmic}
\end{algorithm}

\subsection{Results}
\label{pol_sec}

\begin{figure*}[h]
\centering
\includegraphics[width=1\textwidth]{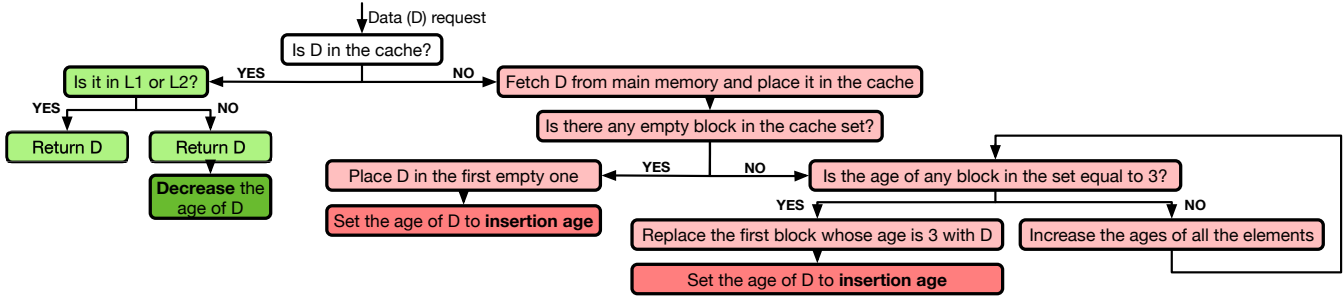}
\caption{Diagram that represents the process of data (D) retrieval whenever the processor makes a request. The blocks with green background represent a cache hit, whereas the blocks with red background represent a cache miss.}
\label{repl_policy}
\end{figure*}

The outcomes of our experiments highlight some differences in the cache architecture of the machines, as also noticed in \cite{disselkoen17}. Traditionally, the cache is divided into slices, each of them containing N sets. The number of slices used to be equal to the number of physical cores a machine has. This is true for the 4th and 5th generation processors. Per contra, the newest ones have as many slices as virtual cores; that is, two times the number of physical cores. Cache sizes are similar, so they also differ in the number of sets per slice.

Since several policies suggest that different sets can perform differently, we have repeated the experiment in Algorithm~\ref{test_pol} for each of the sets in the last level cache. As a result, we have found out that only the i7-4790 and the i3-5010U machines (4th and 5th generation) implement \emph{set dueling} to dynamically select the eviction policy with better performance between two candidate policies. We conducted several further experiments intended for determining which sets implement a fixed policy and which others change their policy based on the number of hits and misses. Locating the sets with a fixed policy is interesting for several reasons: these sets will allow to accurately determine the two different replacement policies and they will allow to favor one policy over the other depending on our interests. This also means that we could monitor one set belonging to the group of followers to determine which policy is currently operating.

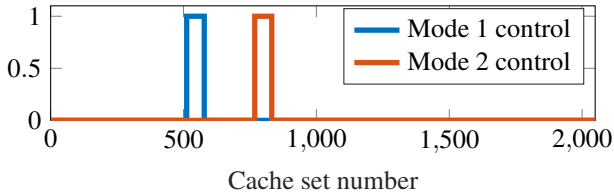
\begin{figure}[tb]
\centering
%
%
\definecolor{mycolor2}{rgb}{0.00000,0.44700,0.74100}%
\definecolor{mycolor1}{rgb}{0.85000,0.32500,0.09800}%
\begin{tikzpicture}

\begin{axis}[%
width=2.85in,
height=0.6in,
at={(0.594in,0.321in)},
scale only axis,
xmin=0,
xmax=2049,
xlabel style={font=\color{white!15!black}},
xlabel={Cache set number},
ymin=0,
ymax=1.1,
axis background/.style={fill=white},
legend style={legend cell align=left, align=left, draw=white!15!black}
]

\addplot [color=mycolor2, line width=2.0pt]
  table[row sep=crcr]{%
1	0\\
512	0\\
512	1\\
578	1\\
578	0\\
2048	0\\
};
\addlegendentry{Mode 1 control}

\addplot [color=mycolor1, line width=2.0pt]
  table[row sep=crcr]{%
1	0\\
768	0\\
768	1\\
832	1\\
832	0\\
2048	0\\
};
\addlegendentry{Mode 2 control}

\end{axis}
\end{tikzpicture}%
\caption{Location of the sets controlling the eviction policy within a slice of 2048 sets. Mode 1 (blue) and mode 2 (red).} \label{control_sets}
\end{figure}

The strategies for locating the sets included different access patterns that we believed would lead to different number of misses. For example, if we access the eviction set in an ordered way, then we access the whole conflicting set, and finally re-access again the eviction set, we will observe different number of misses depending on the policy. Pseudo LRU policies will probably evict all the data in the eviction set after accessing the elements in the conflicting set. Whereas other policies intended for good performance in these situations (burst accesses to memory) will probably incur fewer misses. As a result, we have located two regions composed of 64 cache sets in each slice that control each policy. Figure \ref{control_sets} represents all the sets of a cache slice with the control regions. The region coloured in blue controls the policy 1, and the region coloured in red controls the policy 2. 

\begin{figure}[tb]
\centering
\includegraphics[width=0.5\textwidth,trim={20mm 0 0 0},clip]{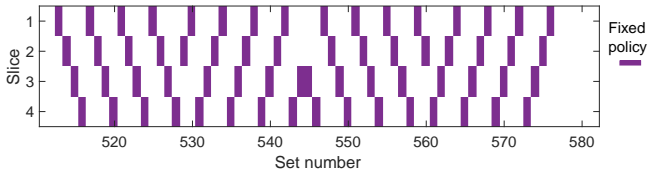}
\caption{Detailed representation of the sets with fixed policy within each of the slices for the i7-4790 machine.} \label{control_sets_d}
\end{figure}

However, not all the sets within the aforementioned regions implement a fixed policy. Particularly, only one of the sets in each slice implements a fixed policy, the corresponding sets in the remaining slices will have a varying policy. This fact was discovered after multiple experiments with different patterns. The sets with fixed policy for each of the slices are depicted in figure \ref{control_sets_d}. To obtain the actual control sets within the slice, it is important to test the sets without order, as it may seem that some sets have a fixed policy and they do not.

The policy we will undercover is the one implemented in the L3 cache. The policies implemented in the L1 and L2 caches can be different. We have been able to uncover a policy that seems to explain the observed evictions. In fact, over 98\% of the evictions have been correctly predicted in all cases \footnote{These results refer to the sets with fixed policy in the machines that implement set dueling. The remaining sets were tested once the two policies were known and we check they followed one of them.}, and it is likely that the errors were due to noise. 

Although we have observed differences between generations and some machines implement set dueling, the decision of which data is going to be evicted is the same in all cases. The replacement policy is always the same and what changes is the insertion policy. Due to space limitations and to avoid creating confusion, we only include here the description of the policies revealed by our experiments as the ones implemented in the Intel processors. Assuming that the policy is named Quad-Age LRU, in the following we refer to ages instead of control bits. Figure \ref{repl_policy} represents the two possible situations whenever the processor requests a piece of data in our processors. If the data is in the LLC, the controller decreases the age of the requested element when giving it to the processor. If there is a cache miss and one element has to be evicted, the replacement policy will select the oldest one. 

Intel processors use two bits to represent the age of the elements in the cache. Consequently, the maximum age is three. In case the reader wonders which block will be replaced in case there are multiple blocks whose age is three, the answer is the first one it finds. The cache behaves somehow like an array of data, and when searching for a block of data placed on it, the controller always starts from the same location, which would be the equivalent to the index 0 in an array.   

As we have already stated, the machines used in our experiments only differ in the insertion age; that is, the value that gets a cache line as age when first loaded in the set or when reloaded after a cache miss. Particularly, the i7-4790 and the i3-5010U machines (4th and 5th generation) that implement \emph{set dueling} insert the elements with age 2 in one of the cases and with age 3 in the other. We denote each of these situations or working modes as mode 1 and mode 2 respectively. The i7-6700K, the i5-7600K and the i7-8650U machines (6th, 7th and 8th generations) always insert the blocks with age 2, which is equivalent to the mode 1 in the oldest machines.

\begin{figure*}[ht]
\centering
\includegraphics[width=1\textwidth]{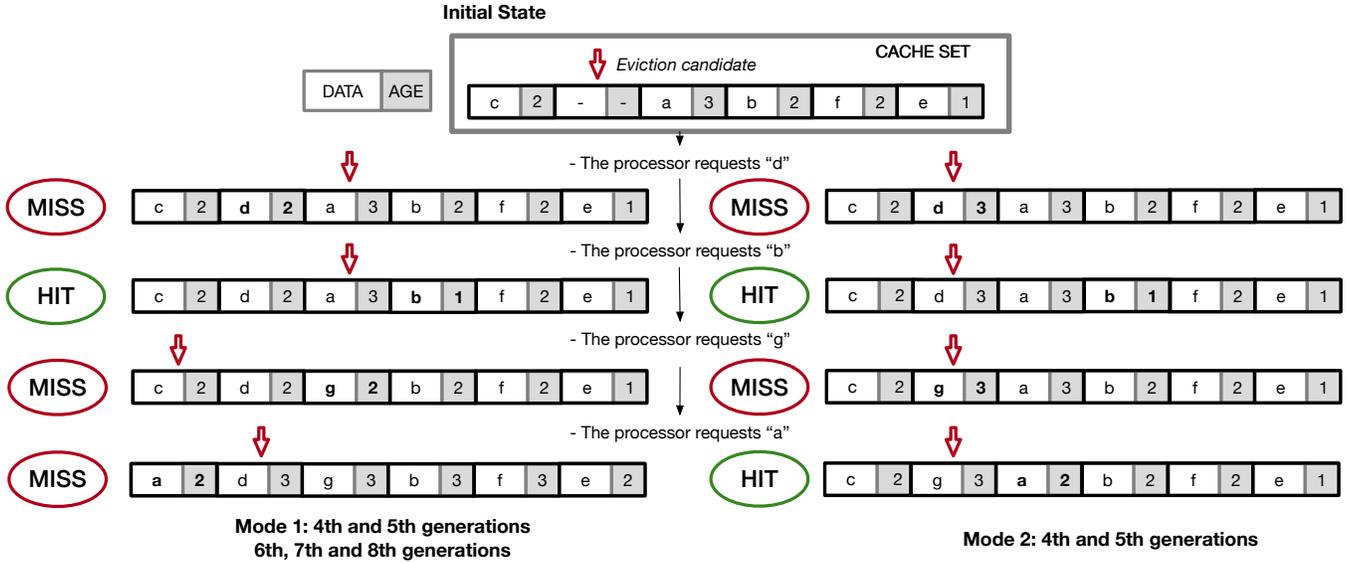}
\caption{Sequence of data accesses in a cache set updating their content and their associated ages for the two observed policies. Mode 1 of the 4th and 5th generations behaves exactly the same as the 6th 7th and 8th generations. The red arrow points the eviction candidate, that is, the data that would be evicted in case of cache miss.}
\label{ex_policy}
\end{figure*}

In order to help the reader to understand how the cache works, figure \ref{ex_policy} shows an example of how the contents of a cache set are updated with each access according to each policy. When the processor requests the line ``d", there is an empty block in the set, so "d" is located in that set and it gets age 2 (Mode 1) or age 3 (Mode 2). In mode 1, the eviction candidate is now ``a" because it is the only one with age 3, whereas in mode 2 the eviction candidate is ``d" as it has age 3 and is on the left of ``a". The processor then requests ``b"  so its age decreases from 2 to 1 in both cases. Accessing ``g" causes a miss. The aforementioned eviction candidates will be replaced with ``g", and its age will be set to 2 or 3 respectively. Eventually, when the processor requests ``a", it will cause a miss in mode 1 (it was evicted on the previous step) and a hit in mode 2, so it will decrease its age.

\section{RELOAD+REFRESH}

If any kind of sharing mechanism is implemented, an attacker knowing the eviction policy can place some data that the victim is likely to use in the cache (the \textit{target}) and in the desired position among the set. Since the position of the blocks and their ages (which in turn depend on the sequence of memory accesses) determine the exact eviction candidate, the attacker can force the \textit{target} to be the eviction candidate. If the victim uses the \textit{target} it will be no longer the eviction candidate, because its age decreases with the access. The attacker can force a miss and check afterwards if the \textit{target} is still in the cache. If it is, the attacker has retrieved the desired information and the victim has retrieved the data from the cache without any cache misses (no attack trace). This is the main idea of the RELOAD+REFRESH attack.

OSs implement mechanisms such as Kernel Same-page Merging (KSM) in Linux \cite{arcangeli09} that improve memory utilization by merging multiple copies of identical memory pages into one. This feature was originally designed for virtual environments where multiple VMs are likely to place the same data in memory, and was later included in the OSs. Although most cloud providers have disabled it, it is still enabled in multiple OSs. When enabled, the attacker needs some reverse engineering to retrieve the address he wants to monitor and he also needs to find an eviction set that maps to the same set as this address. Section \ref{retri} shows how to construct the eviction sets in order to find the one which creates a ``conflict'' with the target address. We follow the procedure in algorithm \ref{conf_ob} replacing the conflicting set candidates with the target address.

\begin{figure}[ht]
\centering
\includegraphics[width=0.5\textwidth]{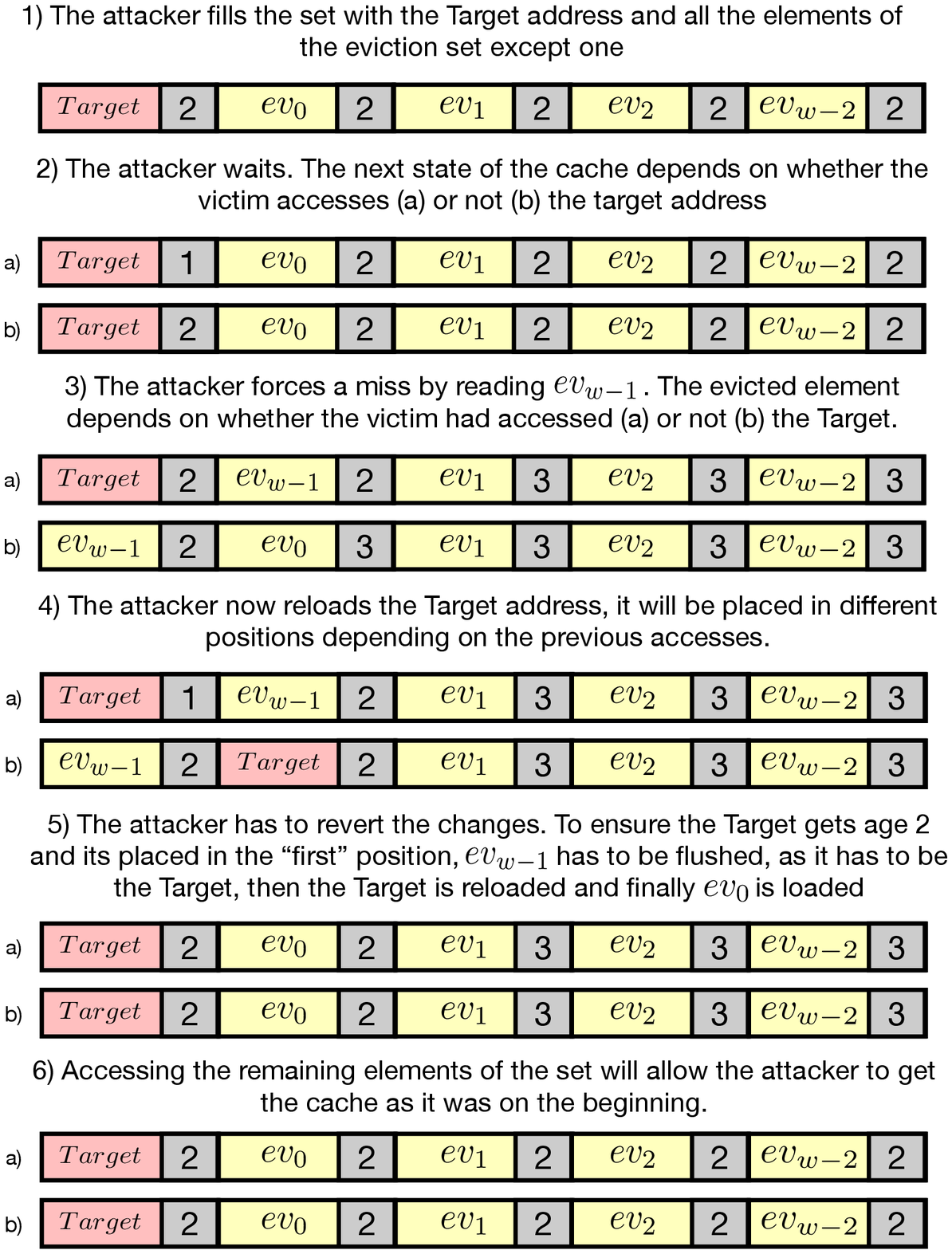}
\caption{Sequence of possible cache set states during the attack for the mode 1 or the newest generations, starting with all elements in the set with age 2.}
\label{at_bas}
\end{figure}

We use figure \ref{at_bas} to depict the stages of the attack and the possible ``states'' of the cache set. The attacker first inserts into the cache the target address and then all the elements in the eviction set except one, which will be used to force an eviction. By the time the attacker has finished filling the cache with data, the target address will be in level 3 cache. The number of ways in low level caches is lower than the number of ways in the L3 cache, and since the L3 cache is inclusive it will remove the target address from the low level caches when loading the last elements of the eviction set. Even if the victim and the attacker are located in the same core, an access of the victim to the target address will update its age, so the attacker would be able to retrieve this information. 

The data is placed in such a way that the target becomes the eviction candidate. The attacker then waits for the victim to access the target. If it does the eviction candidate changes and is now the element inserted in the second place. If it does not, the eviction candidate is still the target address. The attacker reads then the remaining element of the eviction set ($ev_{w-1}$), forcing this way a conflict in the cache set, and the eviction of the candidate. As a consequence, when reading (RELOAD) the target address again, the attacker will know if the victim has used the data (low reload time) or not (high reload time). The state of the cache has to be reverted to the initial one, so all the elements get the same age again (REFRESH). The element $ev_{w-1}$ is forced out of the cache, so it could be used to create a new conflict on the next iteration.

When the cache policy is working in mode 2, each element is inserted with age 3. In this case, steps 1 to 5 are equivalent. However step 6 changes depending on whether the victim is allocated in the same core as the attacker or not. 
When not, the other elements have age 3 and the target is the eviction candidate, so there is no need to refresh the data for the attack. 
On the other hand, when they are on the same core,
the attacker needs to remove the target from the low level cache by refreshing the other elements in the cache. Moreover, the attack could be performed using the low level caches.

The mode 2 policy enables a detectable fast cross core cache attack that does not require shared memory. Once the cache set is filled with the attacker's data, the eviction candidate is now the first element inserted by the attacker. If the victim uses the expected data, the eviction candidate will be replaced. Even if the victim uses the data multiple times, its age will not change, since it will be fetched from the low level caches. Then, after forcing a miss, the attacker only has to access the first element (eviction candidate) to check whether the victim has or has not accessed the target data. Note that with this access the attacker replaces the victim's data (because it became the eviction candidate when loaded) so it is equivalent to the REFRESH. If, on the contrary, the victim does not use the data, the attacker's data will still be in the cache. The attacker will then flush and reload this data to ensure it gets age 3 again.

Algorithms \ref{reload_refresh} and \ref{refresh_step} summarize the steps of the RELOAD+REFRESH attack when the insertion age is two (newest Intel generations or mode 1 in oldest generations). The cache set is filled with the target address plus $W-1$ elements of the eviction set during initialization. 
Then, the attacker waits for the victim to run the code. Later, he performs the RELOAD and REFRESH steps. The RELOAD step gives information about the victim accesses and the REFRESH step gets the set ready to retrieve information from the victim. When initializing the set, we first fill the set, then flush the whole set and finally reload the data again to ensure the insertion order.

In the RELOAD function it is not necessary to flush the Target\_address unless it has not been used by the victim. The same assumption is true for the conflicting address or the element $w-1$ of the eviction set, which would have to be flushed only in that situation. However, to avoid \textit{if} conditions in the code, we have chosen to implement the RELOAD function this way. Low reload times mean the data was used by the victim, whereas high reload times mean it was not.

\renewcommand{\algorithmicrequire}{\textbf{Input:}}
\renewcommand{\algorithmicensure}{\textbf{Output:}}
\begin{algorithm}
\caption{Reload function}\label{reload_refresh}
\begin{algorithmic}[0]
\Require \textbf{Eviction\_set, Target\_address}
\Ensure \textbf{Reload time} 
\Function{RELOAD}{Target\_address,eviction\_set} 
\State \emph{``rdtsc";}
\State \emph{``mfence";}
\State $read(eviction\_set[w-1]);$ \Comment{Forces a miss}
\State \emph{``mfence";}
\State $flush(eviction\_set[w-1]);$
\State \emph{``mfence";}
\State $read(Target\_address);$ 
\State $flush(Target\_address);$
\State \emph{``mfence";}
\State $read(Target\_address);$ \Comment{Reload on first position}
\State \emph{``mfence";}
\State \emph{``rdtsc";}
\State $read(eviction\_set[0]);$
\State \textbf{return} $time\_reload;$
\EndFunction
\end{algorithmic}
\end{algorithm}

\renewcommand{\algorithmicrequire}{\textbf{Input:}}
\renewcommand{\algorithmicensure}{\textbf{Output:}}
\begin{algorithm}
\caption{Refresh function}\label{refresh_step}
\begin{algorithmic}[0]
\Require \textbf{Eviction\_set}
\Ensure \textbf{Refresh time} 
\Function{REFRESH}{Eviction\_set} 
\State volatile unsigned int time;
\State \emph{asm \_\_volatile\_\_(}
\State \quad `` mfence $\setminus$n"
\State \quad `` rdtsc $\setminus$n"
\State \quad `` movl \%\%eax, \%\%esi $\setminus$n"
\State \quad `` movq 8(\%1), \%\%rdi $\setminus$n" \Comment{Eviction\_set[1]}
\State \quad `` movq (\%\%rdi), \%\%rdi $\setminus$n"
\State \quad `` movq (\%\%rdi), \%\%rdi $\setminus$n"
\State \quad `` movq (\%\%rdi), \%\%rdi $\setminus$n"
\State \quad `` movq (\%\%rdi), \%\%rdi $\setminus$n"
\State \quad `` movq (\%\%rdi), \%\%rdi $\setminus$n"
\State \quad `` movq (\%\%rdi), \%\%rdi $\setminus$n"
\State \quad `` movq (\%\%rdi), \%\%rdi $\setminus$n"
\State \quad `` movq (\%\%rdi), \%\%rdi $\setminus$n"
\State \quad `` movq (\%\%rdi), \%\%rdi $\setminus$n"
\State \quad `` movq (\%\%rdi), \%\%rdi $\setminus$n" \Comment{Eviction\_set[w-2]}
\State \quad `` mfence $\setminus$n"
\State \quad `` rdtsc $\setminus$n"
\State \quad `` subl \%\%esi, \%\%eax $\setminus$n" \Comment{Time value on \%eax}
\State \emph{);}
\State \textbf{return} $time\_refresh;$
\EndFunction
\end{algorithmic}
\end{algorithm}

The REFRESH function presented is meant for a 12 way set. Since the target and the first element of the eviction set have been loaded in the RELOAD step, the REFRESH function only has to access the remaining 10 elements of the set. To avoid out of order execution and ensure the order, the elements of the eviction set have to be provided as a linked list (one element contains the address of the following one). Additionally the \textit{refresh time} can be used to detect if any other process is also using that set. 

\subsection{Noise tolerance}

The proposed attack relies on the order in which the elements are inserted into the cache set to both avoid misses on the victim side and to learn information about the data that has been accessed. 
If other processes are running and using data that maps to the same cache slice (introducing noise), the efficiency of the attack can be lessened and also some detection mechanisms can be triggered.

As mentioned before, the refresh step can reveal such situations. Then, the attacker can slightly change the approach. Assuming that only one address is being used by the noise generating process, the attacker can easily handle noise, avoid detection and still gain information about the victim. The trick to deal with noise is placing the target on a different place within the set (the second place in this example). In case somebody else uses any data mapping to that set, the replaced data belongs to the attacker; specifically it is the data placed in first place in the set. When the attacker forces a miss, the eviction candidate will be either the target address (if the victim did not used it) or the element inserted in third place (the victim did use the target data). The attacker can gain information about the victim by reloading the target address and he must begin by refreshing the third element of the eviction set and finish with the first one which will evict the ``noise" from the cache, so the age of all the blocks is 2 again.

\section{Results}

To show the applicability of RELOAD+REFRESH, we have replicated two published attacks:
one against the T-Table implementation of AES and one against the square and multiply exponentiation implementation included in RSA. Although both implementations have been replaced by new ones, we use them for comparison. All the experiments presented in this section have been performed in the intel i5-7600K machine.

\subsection{Attacking AES}

The T-Table implementation used to be a popular software implementation of AES. While still available, this implementation is not the default option when compiling the Openssl library due to its susceptibility to micro architectural attacks. This implementation replaces the \emph{SubBytes}, \emph{ShiftRows} and \emph{MixColumns} operations with table lookups (memory accesses) and XOR operations. Since the accesses to the T-Tables depend on the secret key, an attacker monitoring just one line of each T-Table is able to recover the full AES key.

Our scenario is similar to the one described by Irazoqui et al. in \cite{Irazoqui14}. They focused in retrieving information about the last round of the AES encryption process, where the ciphertext is obtained by performing one XOR operation between an element contained in the tables and the secret key. As the content of the tables is publicly available from the source code, they obtained the secret key xoring it with the ciphertext.

Besides performing the attack against the AES T-Table implementation (Openssl 1.0.1f) using the RELOAD+REFRESH (R+R) technique, we have performed the same attack using the FLUSH+RELOAD (F+R) and PRIME+PROBE (P+P) techniques, to provide a fair comparison regarding to the number of traces required to obtain the key. In order to retrieve the whole key, the attacker has to monitor at least one line of each T-Table. The attacker can monitor from one up to four lines at a time. For this comparison we monitor one table at a time. 

Table \ref{tab:aes} shows the results for each of the approaches. In this scenario the attacker performs one operation, then the victim performs the encryption, and finally the attacker retrieves the information about the victim. That is, victim and attacker do not interfere with each other while doing the different operations. As it can be inferred from the table, our approach performs almost as good as FLUSH+RELOAD, and clearly outperforms PRIME+PROBE.

\begin{table}[htb]
\caption{Mean number of samples required to retrieve each four byte group of the whole AES key when monitoring one line per encryption.}
\label{tab:aes}
\centering
\begin{tabular}{|c|c|c|c|}
\cline{1-4}
\textbf{Attack} & R+R & F+R & P+P \\
\cline{1-4}
\textbf{Samples} & 110000 & 108000 & 220000\\
\cline{1-4}
\end{tabular}
\end{table}

\subsubsection{Detection evaluation}

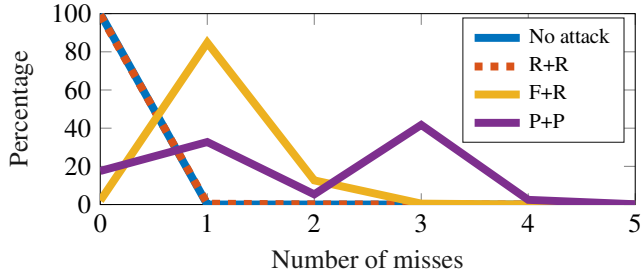
\begin{figure}
\centering
%
%


\definecolor{mycolor1}{rgb}{0.00000,0.44700,0.74100}%
\definecolor{mycolor2}{rgb}{0.85000,0.32500,0.09800}%
\definecolor{mycolor3}{rgb}{0.92900,0.69400,0.12500}%
\definecolor{mycolor4}{rgb}{0.49400,0.18400,0.55600}%
\begin{tikzpicture}

\begin{axis}[%
width=2.8in,
height=1.0in,
at={(1.499in,0.654in)},
scale only axis,
legend style={font=\fontsize{8}{8}\selectfont},
xmin=0,
xmax=5,
xlabel style={font=\color{white!10!black}},
xlabel={Number of misses},
ymin=0,
ymax=100,
ylabel style={font=\color{white!10!black}},
ylabel={Percentage},
axis background/.style={fill=white},
legend style={legend cell align=left, align=left, draw=white!10!black}
]
\addplot [color=mycolor1, line width=3.0pt]
  table[row sep=crcr]{%
0	99.9617\\
1	0.0172\\
2	0.0148\\
3	0.0012\\
4	0.0027\\
5	0.0008\\
};
\addlegendentry{No attack}

\addplot [color=mycolor2, dashed, line width=3.0pt]
  table[row sep=crcr]{%
0	99.4049\\
1	0.4482\\
2	0.1188\\
3	0.0142\\
4	0.0078\\
5	0.0017\\
};
\addlegendentry{R+R}

\addplot [color=mycolor3, line width=3.0pt]
  table[row sep=crcr]{%
0	2.0413\\
1	84.8118\\
2	12.6548\\
3	0.4839\\
4	0.0033\\
5	0.0025\\
};
\addlegendentry{F+R}

\addplot [color=mycolor4, line width=3.0pt]
  table[row sep=crcr]{%
0	17.5758\\
1	32.7437\\
2	5.339\\
3	41.72\\
4	2.4561\\
5	0.1554\\
};
\addlegendentry{P+P}

\end{axis}
\end{tikzpicture}%
\caption{Distribution of the number of misses induced in the victim process by the different attacks, and with no attack. Each includes 1 million samples} \label{aes_dist1}
\end{figure}

RELOAD+REFRESH is able to retrieve an AES key with a negligible impact on the victim process. We compare the number of cache misses the victim suffers per encryption performed, for all the attacks and for normal executions. We use the PAPI software interface \cite{Mucci99papia} to read the counters referred to the victim. PAPI allows us to insert one instruction just before and another one just after the encryption process to read the L3 cache misses counter, which is mainly the information used for cache attack detection \cite{ChiappettaEtAl2016,Zhang2016,Briongos_2018,Kulah2018}. As a result, figure \ref{aes_dist1} shows the distribution of the number of misses the victim sees for each attack and for the normal execution of the encryption.

As figure \ref{aes_dist1} shows, our attack can not be distinguished from the normal performance of the AES encryption process by measuring the number of misses. Note that, when performing the PRIME+PROBE attack against AES, around 20\% of the times the attacker does not cause misses on the victim. When retrieving the information about the victim using this approach, we first perform a PRIME, then wait for the victim to perform an encryption and finally perform a PROBE. During the PRIME stage the elements in the eviction set are accessed from 0 to 11, whereas in the PROBE step they are accessed the opposite way, in a zig-zag pattern. Due to the eviction policy and to the zig-zag pattern some of the elements still reside in low level caches and their ages are not updated. As a consequence, the attack is not able to evict the victim's data from the cache, so the victim does not see any miss and it is likely that the attacker sees a false positive.

\begin{figure}
\centering
%
%
\definecolor{mycolor1}{rgb}{0.00000,0.44700,0.74100}%
\definecolor{mycolor2}{rgb}{0.85000,0.32500,0.09800}%
\definecolor{mycolor3}{rgb}{0.92900,0.69400,0.12500}%
\definecolor{mycolor4}{rgb}{0.49400,0.18400,0.55600}%
\begin{tikzpicture}

\begin{axis}[%
width=2.85in,
height=1.0in,
at={(0.975in,0.287in)},
scale only axis,
xmin=506.619515570934,
xmax=942.854809688581,
xlabel style={font=\color{white!10!black}},
xlabel={Cycles},
ymin=0,
ymax=18,
ylabel style={font=\color{white!10!black}},
ylabel={Percentage},
axis background/.style={fill=white},
legend style={legend cell align=left, align=left, draw=white!10!black}
]
\addplot [color=mycolor1, line width=2.0pt]
  table[row sep=crcr]{%
300	0\\
310	0\\
320	0\\
330	0\\
340	0\\
350	0\\
360	0\\
370	0\\
380	0\\
390	0\\
400	0\\
410	0\\
420	0\\
430	0\\
440	0\\
450	0\\
460	0\\
470	0\\
480	0\\
490	0\\
500	0\\
510	0\\
520	0.0014\\
530	0.0887\\
540	0.6508\\
550	2.7249\\
560	7.1471\\
570	12.351\\
580	15.7958\\
590	16.2045\\
600	14.317\\
610	11.3532\\
620	8.0083\\
630	5.1337\\
640	2.9361\\
650	1.5597\\
660	0.7962\\
670	0.4099\\
680	0.2259\\
690	0.1214\\
700	0.0698\\
710	0.0318\\
720	0.0166\\
730	0.0079\\
740	0.0038\\
750	0.0025\\
760	0.003\\
770	0.0023\\
780	0.0029\\
790	0.0024\\
800	0.0026\\
810	0.0014\\
820	0.0009\\
830	0.0003\\
840	0.0008\\
850	0.0004\\
860	0.0004\\
870	0.0004\\
880	0.0003\\
890	0.0002\\
900	0.0002\\
910	0.0002\\
920	0.0002\\
930	0.0004\\
940	0.0004\\
950	0.0005\\
960	0.0002\\
970	0.0004\\
980	0.0002\\
990	0.0003\\
1000	0.0003\\
1010	0.0004\\
1020	0\\
1030	0\\
1040	0.0002\\
1050	0.0001\\
1060	0.0002\\
1070	0.0001\\
1080	0.0004\\
1090	0.0001\\
1100	0.0002\\
1110	0\\
1120	0\\
1130	0\\
1140	0\\
1150	0\\
1160	0\\
1170	0\\
1180	0\\
1190	0.0001\\
1200	0\\
1210	0.0001\\
1220	0\\
1230	0\\
1240	0\\
1250	0.0002\\
1260	0\\
1270	0\\
1280	0.0001\\
1290	0\\
1300	0\\
1310	0\\
1320	0\\
1330	0\\
1340	0.0001\\
1350	0\\
1360	0\\
1370	0\\
1380	0\\
1390	0\\
1400	0.0001\\
1410	0\\
1420	0\\
1430	0\\
1440	0\\
1450	0\\
1460	0\\
1470	0\\
1480	0\\
1490	0\\
1500	0\\
1510	0\\
1520	0\\
1530	0\\
1540	0\\
1550	0\\
1560	0\\
1570	0\\
1580	0.0001\\
1590	0\\
1600	0\\
1610	0\\
1620	0\\
1630	0\\
1640	0\\
1650	0\\
1660	0\\
1670	0\\
1680	0\\
1690	0\\
1700	0\\
1710	0\\
1720	0\\
1730	0\\
1740	0\\
1750	0\\
1760	0\\
1770	0.0002\\
1780	0\\
1790	0.0001\\
1800	0.0001\\
1810	0\\
1820	0\\
1830	0\\
1840	0.0001\\
1850	0\\
1860	0\\
1870	0\\
1880	0\\
1890	0\\
1900	0\\
1910	0\\
1920	0\\
1930	0\\
1940	0\\
1950	0\\
1960	0\\
1970	0.0001\\
1980	0\\
1990	0\\
2000	0.0173\\
};
\addlegendentry{no attack}

\addplot [color=mycolor2, line width=2.0pt]
  table[row sep=crcr]{%
300	0\\
310	0\\
320	0\\
330	0\\
340	0\\
350	0\\
360	0\\
370	0\\
380	0\\
390	0\\
400	0\\
410	0\\
420	0\\
430	0\\
440	0\\
450	0\\
460	0\\
470	0\\
480	0\\
490	0\\
500	0\\
510	0\\
520	0.0001\\
530	0.0042\\
540	0.0382\\
550	0.2196\\
560	0.7978\\
570	2.1788\\
580	4.6862\\
590	8.2085\\
600	11.9548\\
610	14.4579\\
620	14.8178\\
630	13.1894\\
640	10.4236\\
650	7.537\\
660	4.9012\\
670	2.916\\
680	1.572\\
690	0.8087\\
700	0.395\\
710	0.2237\\
720	0.1208\\
730	0.0724\\
740	0.0444\\
750	0.0401\\
760	0.0487\\
770	0.0525\\
780	0.0584\\
790	0.05\\
800	0.0419\\
810	0.033\\
820	0.0254\\
830	0.0144\\
840	0.0073\\
850	0.0042\\
860	0.0023\\
870	0.0021\\
880	0.0012\\
890	0.0006\\
900	0.0012\\
910	0.0007\\
920	0.0011\\
930	0.0006\\
940	0.0017\\
950	0.001\\
960	0.0012\\
970	0.0007\\
980	0.0009\\
990	0.0005\\
1000	0.0004\\
1010	0.001\\
1020	0.0005\\
1030	0.0005\\
1040	0.0003\\
1050	0.0001\\
1060	0.0001\\
1070	0.0002\\
1080	0.0001\\
1090	0.0002\\
1100	0.0001\\
1110	0.0006\\
1120	0.0002\\
1130	0.0003\\
1140	0.0002\\
1150	0.0003\\
1160	0\\
1170	0.0002\\
1180	0.0002\\
1190	0.0002\\
1200	0.0005\\
1210	0\\
1220	0.0003\\
1230	0.0001\\
1240	0.0002\\
1250	0.0008\\
1260	0.0004\\
1270	0.0002\\
1280	0.0002\\
1290	0.0001\\
1300	0.0001\\
1310	0.0004\\
1320	0\\
1330	0.0002\\
1340	0.0002\\
1350	0.0004\\
1360	0.0002\\
1370	0.0002\\
1380	0.0001\\
1390	0.0001\\
1400	0.0002\\
1410	0.0002\\
1420	0.0003\\
1430	0\\
1440	0\\
1450	0.0001\\
1460	0.0003\\
1470	0.0001\\
1480	0.0003\\
1490	0.0003\\
1500	0\\
1510	0.0001\\
1520	0\\
1530	0.0001\\
1540	0\\
1550	0.0001\\
1560	0.0002\\
1570	0.0001\\
1580	0.0001\\
1590	0.0002\\
1600	0.0004\\
1610	0.0002\\
1620	0.0002\\
1630	0\\
1640	0.0002\\
1650	0.0002\\
1660	0.0001\\
1670	0.0002\\
1680	0\\
1690	0\\
1700	0.0002\\
1710	0.0001\\
1720	0\\
1730	0.0002\\
1740	0.0002\\
1750	0.0005\\
1760	0.0002\\
1770	0.0001\\
1780	0.0001\\
1790	0.0001\\
1800	0.0001\\
1810	0.0002\\
1820	0.0003\\
1830	0.0001\\
1840	0.0001\\
1850	0.0004\\
1860	0\\
1870	0\\
1880	0.0001\\
1890	0.0002\\
1900	0.0002\\
1910	0.0001\\
1920	0.0002\\
1930	0.0002\\
1940	0.0001\\
1950	0.0001\\
1960	0.0003\\
1970	0\\
1980	0.0001\\
1990	0.0002\\
2000	0.0209\\
};
\addlegendentry{R+R}

\addplot [color=mycolor3, line width=2.0pt]
  table[row sep=crcr]{%
300	0\\
310	0\\
320	0\\
330	0\\
340	0\\
350	0\\
360	0\\
370	0\\
380	0\\
390	0\\
400	0\\
410	0\\
420	0\\
430	0\\
440	0\\
450	0\\
460	0\\
470	0\\
480	0\\
490	0\\
500	0\\
510	0\\
520	0\\
530	0.0003\\
540	0.0033\\
550	0.0335\\
560	0.1911\\
570	0.6498\\
580	1.5003\\
590	2.4911\\
600	3.2654\\
610	3.5514\\
620	3.3955\\
630	3.0679\\
640	2.6014\\
650	2.1402\\
660	1.8287\\
670	1.6418\\
680	1.5943\\
690	1.5574\\
700	1.5927\\
710	1.7359\\
720	1.9538\\
730	2.4694\\
740	3.5441\\
750	5.0767\\
760	6.7606\\
770	7.9701\\
780	8.1369\\
790	7.5357\\
800	6.1277\\
810	4.5521\\
820	3.1361\\
830	1.9615\\
840	1.1711\\
850	0.6666\\
860	0.374\\
870	0.2354\\
880	0.1493\\
890	0.113\\
900	0.0831\\
910	0.0692\\
920	0.059\\
930	0.0522\\
940	0.0487\\
950	0.0451\\
960	0.0421\\
970	0.0476\\
980	0.0413\\
990	0.0397\\
1000	0.0393\\
1010	0.0394\\
1020	0.0404\\
1030	0.0375\\
1040	0.0383\\
1050	0.0402\\
1060	0.0387\\
1070	0.0406\\
1080	0.0375\\
1090	0.0409\\
1100	0.0414\\
1110	0.044\\
1120	0.0452\\
1130	0.0416\\
1140	0.0416\\
1150	0.0391\\
1160	0.0422\\
1170	0.0414\\
1180	0.0431\\
1190	0.042\\
1200	0.0451\\
1210	0.0412\\
1220	0.0403\\
1230	0.038\\
1240	0.0453\\
1250	0.0393\\
1260	0.0427\\
1270	0.0399\\
1280	0.039\\
1290	0.0399\\
1300	0.0434\\
1310	0.0422\\
1320	0.0426\\
1330	0.0413\\
1340	0.0439\\
1350	0.0433\\
1360	0.041\\
1370	0.043\\
1380	0.0415\\
1390	0.0407\\
1400	0.0459\\
1410	0.0413\\
1420	0.0388\\
1430	0.0409\\
1440	0.04\\
1450	0.0414\\
1460	0.0462\\
1470	0.0438\\
1480	0.0426\\
1490	0.0437\\
1500	0.0402\\
1510	0.0427\\
1520	0.0441\\
1530	0.0421\\
1540	0.0431\\
1550	0.0418\\
1560	0.0418\\
1570	0.0438\\
1580	0.0436\\
1590	0.0454\\
1600	0.0432\\
1610	0.0428\\
1620	0.0414\\
1630	0.0418\\
1640	0.0433\\
1650	0.0454\\
1660	0.0453\\
1670	0.0422\\
1680	0.0438\\
1690	0.0384\\
1700	0.0431\\
1710	0.0452\\
1720	0.0424\\
1730	0.0437\\
1740	0.0474\\
1750	0.0412\\
1760	0.0429\\
1770	0.0378\\
1780	0.0435\\
1790	0.0417\\
1800	0.0423\\
1810	0.0423\\
1820	0.0385\\
1830	0.0381\\
1840	0.0432\\
1850	0.0417\\
1860	0.0439\\
1870	0.0427\\
1880	0.0409\\
1890	0.0405\\
1900	0.0398\\
1910	0.0387\\
1920	0.0406\\
1930	0.0381\\
1940	0.0415\\
1950	0.0386\\
1960	0.0372\\
1970	0.0386\\
1980	0.0376\\
1990	0.0369\\
2000	0.5365\\
};
\addlegendentry{F+R}

\addplot [color=mycolor4, line width=2.0pt]
  table[row sep=crcr]{%
300	0\\
310	0\\
320	0\\
330	0\\
340	0\\
350	0\\
360	0\\
370	0\\
380	0\\
390	0\\
400	0\\
410	0\\
420	0\\
430	0\\
440	0\\
450	0\\
460	0\\
470	0\\
480	0\\
490	0\\
500	0\\
510	0\\
520	0\\
530	0.0137\\
540	0.0919\\
550	0.4495\\
560	1.3426\\
570	2.7263\\
580	3.9786\\
590	4.5584\\
600	4.4638\\
610	3.8164\\
620	2.9926\\
630	2.2127\\
640	1.551\\
650	1.0788\\
660	0.8155\\
670	0.6927\\
680	0.6482\\
690	0.6484\\
700	0.7041\\
710	0.8536\\
720	1.2343\\
730	2.0794\\
740	3.6348\\
750	5.8254\\
760	7.9348\\
770	8.9189\\
780	8.7261\\
790	7.4954\\
800	5.7404\\
810	4.0926\\
820	2.6369\\
830	1.5547\\
840	0.8715\\
850	0.4818\\
860	0.2696\\
870	0.1687\\
880	0.112\\
890	0.0822\\
900	0.0643\\
910	0.0503\\
920	0.0462\\
930	0.0421\\
940	0.038\\
950	0.0348\\
960	0.0329\\
970	0.0312\\
980	0.0313\\
990	0.0313\\
1000	0.0289\\
1010	0.0313\\
1020	0.0305\\
1030	0.0286\\
1040	0.0271\\
1050	0.0295\\
1060	0.0305\\
1070	0.0277\\
1080	0.0282\\
1090	0.0275\\
1100	0.035\\
1110	0.0317\\
1120	0.0293\\
1130	0.0279\\
1140	0.0315\\
1150	0.0296\\
1160	0.0316\\
1170	0.0298\\
1180	0.0297\\
1190	0.0307\\
1200	0.0312\\
1210	0.0316\\
1220	0.0325\\
1230	0.0314\\
1240	0.031\\
1250	0.0311\\
1260	0.0322\\
1270	0.0352\\
1280	0.0308\\
1290	0.0328\\
1300	0.0325\\
1310	0.0314\\
1320	0.0331\\
1330	0.0342\\
1340	0.0318\\
1350	0.0333\\
1360	0.0353\\
1370	0.0368\\
1380	0.0366\\
1390	0.0349\\
1400	0.0368\\
1410	0.0347\\
1420	0.0377\\
1430	0.0309\\
1440	0.0381\\
1450	0.0317\\
1460	0.0383\\
1470	0.0355\\
1480	0.0334\\
1490	0.037\\
1500	0.0391\\
1510	0.0349\\
1520	0.0371\\
1530	0.0367\\
1540	0.0378\\
1550	0.0358\\
1560	0.0376\\
1570	0.037\\
1580	0.0378\\
1590	0.0419\\
1600	0.0358\\
1610	0.0375\\
1620	0.0421\\
1630	0.0389\\
1640	0.0379\\
1650	0.0371\\
1660	0.037\\
1670	0.0386\\
1680	0.0395\\
1690	0.0399\\
1700	0.0407\\
1710	0.0375\\
1720	0.0379\\
1730	0.0413\\
1740	0.0415\\
1750	0.0406\\
1760	0.0392\\
1770	0.0421\\
1780	0.0436\\
1790	0.0386\\
1800	0.0394\\
1810	0.0416\\
1820	0.041\\
1830	0.0359\\
1840	0.039\\
1850	0.0389\\
1860	0.0429\\
1870	0.0383\\
1880	0.039\\
1890	0.0397\\
1900	0.0343\\
1910	0.042\\
1920	0.0394\\
1930	0.0394\\
1940	0.0374\\
1950	0.0355\\
1960	0.0354\\
1970	0.0374\\
1980	0.0354\\
1990	0.0377\\
2000	0.5563\\
};
\addlegendentry{P+P}

\end{axis}
\end{tikzpicture}%
\caption{Distribution of the encryption times in different situations. Each distribution includes 1 million samples.}

\label{aes_dist_time}
\end{figure}
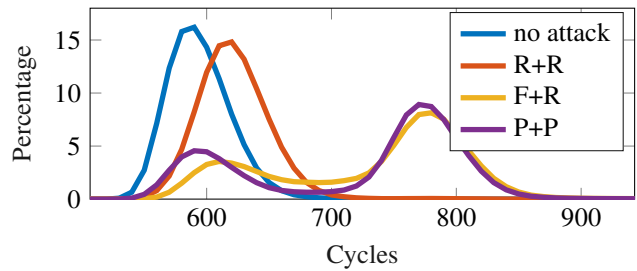

Additionally, we use the \emph{rdtsc} instruction to measure the time it takes to complete each encryption and show the results in figure \ref{aes_dist_time}. The differences observed in figure \ref{aes_dist_time} between the normal encryption and the RELOAD+REFRESH approach are not significant if we compare them with the other attacks. The mean encryption time when there is no attack is 595 cycles, whereas it increases up to 623 cycles when attacked with the RELOAD+REFRESH technique. This time difference exists because, when suffering the RELOAD+REFRESH attack, the victim has to load the data from the L3 cache instead of loading it from the L1 or L2 caches.

\subsection{Attacking RSA}
RSA is the most widely used public key crypto system used for data encryption as well as for digital signatures. Its security is based on the practical difficulty of the factorization of the product of two large prime numbers. RSA involves a public key (used for encryption) and a private key (used for decryption). There are many algorithms suitable for computing the modular exponentiation required for both encryption and decryption. In this work we focus in the \textit{square and multiply} exponentiation algorithm \cite{Gordon1998} as Yarom et al. did in \cite{YaromEtAl2014}. 

Square and multiply computes \textit{x} = \textit{b}$^e$ mod \textit{m} as a sequence of Square and Multiply operations (followed by a Modulo Reduce) that depend on the bits of the exponent e. If the bit happens to be a 1, then the Square-Multiply-Reduce sequence of operations is executed. On the other hand, if the bit is a 0, only the Square-Reduce operations are executed. As a consequence, retrieving the sequence of operations executed means recovering the exponent; that is, the key. 

As a difference with the attack against AES, we monitor instructions instead of data. Additionally, an attack against RSA needs to have enough time resolution to correctly retrieve the sequence of operations. As we did with AES, we performed the attack using our stealthy technique as well as the FLUSH+RELOAD and PRIME+PROBE techniques.

In this work we use the \emph{libgcrypt} version 1.5.0, which includes the aforementioned square and multiply implementation. The key length in our experiments is 2048 bits. When attacking RSA it is possible to monitor all the functions implied in the exponentiation or just one. When monitoring all the instructions the attacker is able to reconstruct the sequence of observations, when monitoring only one instruction the attacker has to use the differences of times between occurrences of the monitored event to retrieve the key. Since our purpose is to compare the ability of our approach to obtain information about the victim, we only monitor the multiply operation. All the attacks were configured to obtain a sample with the same time resolution (3000 cycles), to ensure the differences in accuracy are due to the attack technique. 

\begin{figure*}
\centering
\input{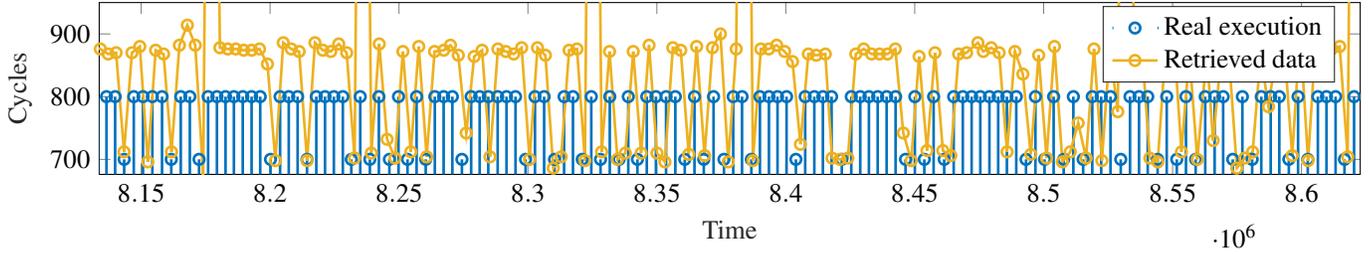}
\caption{Example of a retrieved trace referred to an execution of a RSA decryption. The blue bars represent the real execution of squares (points equal to 800) and multiplies (700). The yellow line represents the information retrieved, low reload times mean detection of the multiply execution.} \label{rsa_ret}
\end{figure*}

Figure \ref{rsa_ret} shows an example of part of a retrieved trace using the RELOAD+REFRESH approach. The trace corresponding to the real sequence of squares and multiplies is represented as blue bars with different values, 800 means a square was executed and 700 it was a multiply. Since the \textit{timestamp} is collected after each exponentiation operation has finished, and the \textit{timestamp} of the attack samples after the reload operation, it may seem that there is some misalignment between traces.

The results of our experiments are summarized in table \ref{tab:rsa}. The accuracy is given as the number of multiplies correctly detected divided by the number of multiplies really executed during the RSA decryption. The percentage of false positives is obtained as the total number a multiply was detected minus the number of correctly detected operations, divided by the total number of detected multiplies:

\[FP=\dfrac{Mul_{detected}-Mul_{correct}}{Mul_{detected}}\]

Again, the RELOAD+REFRESH performance is similar to  FLUSH+RELOAD. The increase in the number of false positives for the PRIME+PROBE approach is most likely due to the way each sample is obtained. In this case, we measure the prime time, then wait for around 3000 cycles, and then measure the probe time, and so on and so forth. Each prime or probe operation gives the time measurement required for detecting the accesses. Taking into the account the eviction policy, it is expected that sometimes the victim data still resides in the cache after a prime or probe step.

\begin{table}[htb]
\caption{Percentage of samples correctly retrieved and false positives generated by each approach when attacking RSA.}
\label{tab:rsa}
\centering
\begin{tabular*}{0.9\linewidth}{|c|c|c|c|}
\cline{1-4}
\textbf{Attack} & R+R & F+R & P+P \\
\cline{1-4}
\textbf{True positives} & 94.9\% & 97.25 \% & 96.16\% \\
\cline{1-4}
\textbf{False positives} & 38.75\% & 38.86\% & 57.25\% \\
\cline{1-4}
\end{tabular*}
\end{table}

\subsubsection{Detection evaluation}

\begin{figure}
\centering
%
%
\definecolor{mycolor1}{rgb}{0.00000,0.44700,0.74100}%
\definecolor{mycolor2}{rgb}{0.85000,0.32500,0.09800}%
\definecolor{mycolor3}{rgb}{0.92900,0.69400,0.12500}%
\definecolor{mycolor4}{rgb}{0.49400,0.18400,0.55600}%
\begin{tikzpicture}

\begin{axis}[%
width=2.7in,
height=0.8in,
at={(0.977in,0.424in)},
scale only axis,
xmin=3997.53866538439,
xmax=9441.86521419925,
xlabel style={font=\color{white!10!black}},
xlabel={Number of L3 cache misses},
ymin=0,
ymax=300,
ylabel style={font=\color{white!10!black}},
ylabel={Amount of samples},
axis background/.style={fill=white},
legend style={at={(0.744,0.40)}, anchor=south west, legend cell align=left, align=left, draw=white!10!black}
]
\addplot [color=mycolor1, line width=2.0pt]
  table[row sep=crcr]{%
1000	0\\
1050	0\\
1100	0\\
1150	0\\
1200	0\\
1250	0\\
1300	0\\
1350	0\\
1400	0\\
1450	0\\
1500	0\\
1550	0\\
1600	0\\
1650	0\\
1700	0\\
1750	0\\
1800	0\\
1850	0\\
1900	0\\
1950	0\\
2000	0\\
2050	0\\
2100	0\\
2150	0\\
2200	0\\
2250	0\\
2300	0\\
2350	0\\
2400	0\\
2450	0\\
2500	0\\
2550	0\\
2600	0\\
2650	0\\
2700	0\\
2750	0\\
2800	0\\
2850	0\\
2900	0\\
2950	0\\
3000	0\\
3050	0\\
3100	0\\
3150	0\\
3200	0\\
3250	0\\
3300	0\\
3350	0\\
3400	0\\
3450	0\\
3500	0\\
3550	0\\
3600	0\\
3650	0\\
3700	0\\
3750	0\\
3800	0\\
3850	0\\
3900	0\\
3950	0\\
4000	0\\
4050	0\\
4100	0\\
4150	0\\
4200	0\\
4250	0\\
4300	1\\
4350	3\\
4400	2\\
4450	1\\
4500	2\\
4550	1\\
4600	2\\
4650	1\\
4700	1\\
4750	0\\
4800	0\\
4850	0\\
4900	0\\
4950	0\\
5000	0\\
5050	1\\
5100	1\\
5150	0\\
5200	0\\
5250	1\\
5300	3\\
5350	0\\
5400	2\\
5450	0\\
5500	0\\
5550	0\\
5600	0\\
5650	0\\
5700	1\\
5750	1\\
5800	2\\
5850	0\\
5900	2\\
5950	1\\
6000	2\\
6050	0\\
6100	3\\
6150	14\\
6200	87\\
6250	187\\
6300	242\\
6350	226\\
6400	145\\
6450	45\\
6500	16\\
6550	1\\
6600	1\\
6650	0\\
6700	0\\
6750	0\\
6800	0\\
6850	0\\
6900	0\\
6950	0\\
7000	0\\
7050	0\\
7100	0\\
7150	0\\
7200	0\\
7250	0\\
7300	0\\
7350	0\\
7400	0\\
7450	0\\
7500	0\\
7550	0\\
7600	0\\
7650	0\\
7700	0\\
7750	0\\
7800	1\\
7850	0\\
7900	0\\
7950	0\\
8000	0\\
8050	1\\
8100	0\\
8150	0\\
8200	0\\
8250	0\\
8300	0\\
8350	0\\
8400	0\\
8450	0\\
8500	0\\
8550	0\\
8600	0\\
8650	0\\
8700	0\\
8750	0\\
8800	0\\
8850	0\\
8900	0\\
8950	0\\
9000	0\\
9050	0\\
9100	0\\
9150	0\\
9200	0\\
9250	0\\
9300	0\\
9350	0\\
9400	0\\
9450	0\\
9500	0\\
9550	0\\
9600	0\\
9650	0\\
9700	0\\
9750	0\\
9800	0\\
9850	0\\
9900	0\\
9950	0\\
10000	0\\
};
\addlegendentry{no attack}

\addplot [color=mycolor2, line width=2.0pt]
  table[row sep=crcr]{%
1000	0\\
1050	0\\
1100	0\\
1150	0\\
1200	0\\
1250	0\\
1300	0\\
1350	0\\
1400	0\\
1450	0\\
1500	0\\
1550	0\\
1600	0\\
1650	0\\
1700	0\\
1750	0\\
1800	0\\
1850	0\\
1900	0\\
1950	0\\
2000	0\\
2050	0\\
2100	0\\
2150	0\\
2200	0\\
2250	0\\
2300	0\\
2350	0\\
2400	0\\
2450	0\\
2500	0\\
2550	0\\
2600	0\\
2650	0\\
2700	0\\
2750	0\\
2800	0\\
2850	0\\
2900	0\\
2950	0\\
3000	0\\
3050	0\\
3100	0\\
3150	0\\
3200	0\\
3250	0\\
3300	0\\
3350	0\\
3400	0\\
3450	0\\
3500	0\\
3550	0\\
3600	0\\
3650	0\\
3700	0\\
3750	0\\
3800	0\\
3850	0\\
3900	0\\
3950	0\\
4000	0\\
4050	0\\
4100	0\\
4150	0\\
4200	0\\
4250	0\\
4300	0\\
4350	0\\
4400	0\\
4450	0\\
4500	0\\
4550	0\\
4600	0\\
4650	0\\
4700	0\\
4750	2\\
4800	9\\
4850	24\\
4900	49\\
4950	102\\
5000	139\\
5050	153\\
5100	112\\
5150	111\\
5200	57\\
5250	48\\
5300	26\\
5350	14\\
5400	24\\
5450	22\\
5500	12\\
5550	19\\
5600	18\\
5650	13\\
5700	6\\
5750	4\\
5800	8\\
5850	6\\
5900	2\\
5950	1\\
6000	3\\
6050	2\\
6100	0\\
6150	1\\
6200	2\\
6250	0\\
6300	0\\
6350	1\\
6400	0\\
6450	0\\
6500	0\\
6550	0\\
6600	1\\
6650	0\\
6700	0\\
6750	1\\
6800	1\\
6850	0\\
6900	1\\
6950	1\\
7000	0\\
7050	0\\
7100	0\\
7150	1\\
7200	0\\
7250	0\\
7300	0\\
7350	0\\
7400	0\\
7450	1\\
7500	0\\
7550	0\\
7600	0\\
7650	0\\
7700	0\\
7750	0\\
7800	0\\
7850	1\\
7900	0\\
7950	0\\
8000	0\\
8050	0\\
8100	0\\
8150	0\\
8200	0\\
8250	0\\
8300	0\\
8350	0\\
8400	0\\
8450	0\\
8500	0\\
8550	1\\
8600	0\\
8650	0\\
8700	0\\
8750	0\\
8800	0\\
8850	0\\
8900	0\\
8950	1\\
9000	0\\
9050	0\\
9100	0\\
9150	0\\
9200	0\\
9250	0\\
9300	0\\
9350	0\\
9400	0\\
9450	0\\
9500	0\\
9550	0\\
9600	0\\
9650	0\\
9700	0\\
9750	0\\
9800	0\\
9850	0\\
9900	0\\
9950	0\\
10000	0\\
};
\addlegendentry{R+R}

\addplot [color=mycolor3, line width=2.0pt]
  table[row sep=crcr]{%
1000	0\\
1050	0\\
1100	0\\
1150	0\\
1200	0\\
1250	0\\
1300	0\\
1350	0\\
1400	0\\
1450	0\\
1500	0\\
1550	0\\
1600	0\\
1650	0\\
1700	0\\
1750	0\\
1800	0\\
1850	0\\
1900	0\\
1950	0\\
2000	0\\
2050	0\\
2100	0\\
2150	0\\
2200	0\\
2250	0\\
2300	0\\
2350	0\\
2400	0\\
2450	0\\
2500	0\\
2550	0\\
2600	0\\
2650	0\\
2700	0\\
2750	0\\
2800	0\\
2850	0\\
2900	0\\
2950	0\\
3000	0\\
3050	0\\
3100	0\\
3150	0\\
3200	0\\
3250	0\\
3300	0\\
3350	0\\
3400	0\\
3450	0\\
3500	0\\
3550	0\\
3600	0\\
3650	0\\
3700	0\\
3750	0\\
3800	0\\
3850	0\\
3900	0\\
3950	0\\
4000	0\\
4050	0\\
4100	0\\
4150	0\\
4200	0\\
4250	0\\
4300	0\\
4350	0\\
4400	0\\
4450	0\\
4500	0\\
4550	0\\
4600	0\\
4650	0\\
4700	0\\
4750	0\\
4800	0\\
4850	0\\
4900	0\\
4950	0\\
5000	0\\
5050	0\\
5100	0\\
5150	0\\
5200	0\\
5250	0\\
5300	0\\
5350	0\\
5400	0\\
5450	0\\
5500	0\\
5550	0\\
5600	0\\
5650	0\\
5700	0\\
5750	0\\
5800	0\\
5850	0\\
5900	0\\
5950	0\\
6000	0\\
6050	0\\
6100	0\\
6150	0\\
6200	0\\
6250	0\\
6300	0\\
6350	0\\
6400	0\\
6450	0\\
6500	0\\
6550	0\\
6600	0\\
6650	0\\
6700	0\\
6750	0\\
6800	0\\
6850	0\\
6900	0\\
6950	0\\
7000	0\\
7050	0\\
7100	0\\
7150	0\\
7200	0\\
7250	0\\
7300	0\\
7350	0\\
7400	0\\
7450	0\\
7500	0\\
7550	0\\
7600	0\\
7650	0\\
7700	0\\
7750	0\\
7800	0\\
7850	0\\
7900	0\\
7950	0\\
8000	0\\
8050	1\\
8100	3\\
8150	6\\
8200	6\\
8250	22\\
8300	41\\
8350	54\\
8400	85\\
8450	101\\
8500	114\\
8550	108\\
8600	127\\
8650	92\\
8700	73\\
8750	64\\
8800	38\\
8850	24\\
8900	6\\
8950	5\\
9000	3\\
9050	6\\
9100	3\\
9150	3\\
9200	1\\
9250	1\\
9300	1\\
9350	1\\
9400	0\\
9450	0\\
9500	1\\
9550	1\\
9600	1\\
9650	1\\
9700	1\\
9750	0\\
9800	1\\
9850	1\\
9900	0\\
9950	0\\
10000	4\\
};
\addlegendentry{F+R}

\addplot [color=mycolor4, line width=2.0pt]
  table[row sep=crcr]{%
1000	0\\
1050	0\\
1100	0\\
1150	0\\
1200	0\\
1250	0\\
1300	0\\
1350	0\\
1400	0\\
1450	0\\
1500	0\\
1550	0\\
1600	0\\
1650	0\\
1700	0\\
1750	0\\
1800	0\\
1850	0\\
1900	0\\
1950	0\\
2000	0\\
2050	0\\
2100	0\\
2150	5\\
2200	6\\
2250	5\\
2300	12\\
2350	17\\
2400	20\\
2450	15\\
2500	16\\
2550	16\\
2600	11\\
2650	14\\
2700	21\\
2750	17\\
2800	20\\
2850	25\\
2900	26\\
2950	21\\
3000	21\\
3050	15\\
3100	22\\
3150	18\\
3200	7\\
3250	15\\
3300	14\\
3350	6\\
3400	11\\
3450	6\\
3500	5\\
3550	6\\
3600	2\\
3650	2\\
3700	1\\
3750	1\\
3800	1\\
3850	1\\
3900	4\\
3950	0\\
4000	1\\
4050	1\\
4100	1\\
4150	1\\
4200	0\\
4250	1\\
4300	0\\
4350	0\\
4400	2\\
4450	1\\
4500	0\\
4550	0\\
4600	1\\
4650	0\\
4700	0\\
4750	0\\
4800	1\\
4850	0\\
4900	1\\
4950	0\\
5000	0\\
5050	0\\
5100	0\\
5150	0\\
5200	0\\
5250	0\\
5300	0\\
5350	0\\
5400	0\\
5450	1\\
5500	4\\
5550	10\\
5600	23\\
5650	34\\
5700	79\\
5750	121\\
5800	108\\
5850	75\\
5900	47\\
5950	17\\
6000	8\\
6050	1\\
6100	5\\
6150	3\\
6200	4\\
6250	1\\
6300	2\\
6350	3\\
6400	5\\
6450	3\\
6500	1\\
6550	0\\
6600	0\\
6650	1\\
6700	0\\
6750	0\\
6800	0\\
6850	2\\
6900	0\\
6950	1\\
7000	0\\
7050	0\\
7100	0\\
7150	0\\
7200	0\\
7250	0\\
7300	0\\
7350	0\\
7400	0\\
7450	1\\
7500	0\\
7550	0\\
7600	0\\
7650	0\\
7700	0\\
7750	0\\
7800	0\\
7850	1\\
7900	0\\
7950	0\\
8000	0\\
8050	1\\
8100	1\\
8150	0\\
8200	0\\
8250	0\\
8300	0\\
8350	0\\
8400	0\\
8450	0\\
8500	0\\
8550	0\\
8600	0\\
8650	0\\
8700	0\\
8750	0\\
8800	0\\
8850	0\\
8900	0\\
8950	0\\
9000	0\\
9050	0\\
9100	0\\
9150	0\\
9200	0\\
9250	0\\
9300	0\\
9350	0\\
9400	0\\
9450	0\\
9500	0\\
9550	0\\
9600	0\\
9650	0\\
9700	0\\
9750	0\\
9800	0\\
9850	0\\
9900	0\\
9950	0\\
10000	1\\
};
\addlegendentry{P+P}

\end{axis}
\end{tikzpicture}%
\caption{Distribution of the number of misses induced in the victim process by the different attacks, and with no attack.} \label{rsa_dist}
\end{figure}
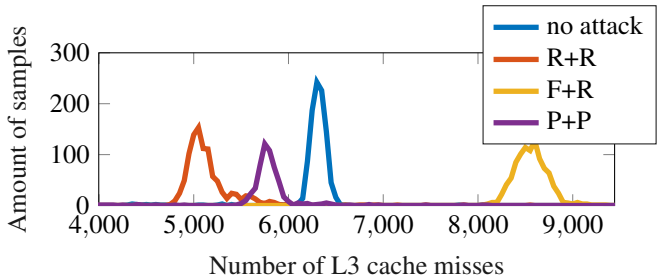

We have monitored the number of cache misses detected when executing a RSA decryption. Figure \ref{rsa_dist} includes the resulting distributions for 1000 samples without attack and applying the considered attacks. The number of total misses increases compared to the no-attack case for the FLUSH+RELOAD attack. However, it decreases for both the RELOAD+REFRESH and PRIME+PROBE attacks, being RELOAD+REFRESH the approach causing the minimum amount of misses. This result is explained due to the monitoring process, to the PRIME+PROBE attack that does not evict the data from the cache, and to the fact that in the RSA decryption most misses happen at the beginning of the process. These results prove that the total amount of misses per decryption is not indicative of an attack going on for RSA.

\begin{figure}[htb]
\centering
%
%
\definecolor{mycolor1}{rgb}{0.00000,0.44700,0.74100}%
\definecolor{mycolor2}{rgb}{0.85000,0.32500,0.09800}%
\definecolor{mycolor3}{rgb}{0.92900,0.69400,0.12500}%
\definecolor{mycolor4}{rgb}{0.49400,0.18400,0.55600}%
\begin{tikzpicture}

\begin{axis}[%
width=2.75in,
height=0.8in,
at={(1.294in,0.424in)},
scale only axis,
xmin=39.3710643733038,
xmax=69.4868450442565,
xlabel style={font=\color{white!15!black}},
xlabel={Sample number},
ymin=-0.0832864274774163,
ymax=151.512325469601,
ylabel style={font=\color{white!15!black}},
ylabel={LLC misses},
axis background/.style={fill=white},
legend style={at={(0.72,0.428)}, anchor=south west, legend cell align=left, align=left, draw=white!15!black}
]
\addplot [color=mycolor1, line width=2.0pt]
  table[row sep=crcr]{%
1	2583\\
2	0\\
3	14\\
4	0\\
5	0\\
6	0\\
7	0\\
8	0\\
9	0\\
10	0\\
11	0\\
12	0\\
13	0\\
14	0\\
15	0\\
16	0\\
17	161\\
18	294\\
19	29\\
20	6\\
21	100\\
22	76\\
23	32\\
24	6\\
25	0\\
26	0\\
27	52\\
28	80\\
29	60\\
30	22\\
31	0\\
32	0\\
33	0\\
34	44\\
35	72\\
36	64\\
37	26\\
38	0\\
39	19\\
40	0\\
41	6\\
42	0\\
43	0\\
44	0\\
45	0\\
46	0\\
47	0\\
48	0\\
49	0\\
50	0\\
51	0\\
52	0\\
53	0\\
54	0\\
55	0\\
56	0\\
57	0\\
58	0\\
59	0\\
60	0\\
61	0\\
62	0\\
63	0\\
64	0\\
65	0\\
66	0\\
67	335\\
};
\addlegendentry{no attack}

\addplot [color=mycolor2, line width=2.0pt]
  table[row sep=crcr]{%
1	1276\\
2	0\\
3	2\\
4	0\\
5	0\\
6	0\\
7	0\\
8	0\\
9	0\\
10	0\\
11	0\\
12	0\\
13	0\\
14	0\\
15	0\\
16	278\\
17	125\\
18	34\\
19	38\\
20	89\\
21	84\\
22	66\\
23	32\\
24	16\\
25	8\\
26	12\\
27	55\\
28	68\\
29	43\\
30	16\\
31	16\\
32	4\\
33	8\\
34	78\\
35	76\\
36	51\\
37	31\\
38	53\\
39	8\\
40	6\\
41	56\\
42	6\\
43	16\\
44	3\\
45	12\\
46	8\\
47	6\\
48	12\\
49	10\\
50	2\\
51	0\\
52	8\\
53	6\\
54	12\\
55	4\\
56	14\\
57	1\\
58	3\\
59	8\\
60	4\\
61	14\\
62	6\\
63	8\\
64	2\\
65	8\\
66	17\\
67	201\\
};
\addlegendentry{R+R}

\addplot [color=mycolor3, line width=2.0pt]
  table[row sep=crcr]{%
1	2656\\
2	0\\
3	0\\
4	0\\
5	0\\
6	0\\
7	0\\
8	0\\
9	0\\
10	0\\
11	0\\
12	0\\
13	0\\
14	0\\
15	1\\
16	0\\
17	385\\
18	270\\
19	120\\
20	168\\
21	164\\
22	156\\
23	112\\
24	54\\
25	92\\
26	100\\
27	140\\
28	162\\
29	158\\
30	132\\
31	106\\
32	102\\
33	94\\
34	142\\
35	172\\
36	152\\
37	108\\
38	76\\
39	94\\
40	29\\
41	8\\
42	125\\
43	98\\
44	94\\
45	82\\
46	80\\
47	94\\
48	106\\
49	80\\
50	90\\
51	100\\
52	80\\
53	84\\
54	66\\
55	102\\
56	96\\
57	90\\
58	108\\
59	96\\
60	102\\
61	84\\
62	82\\
63	94\\
64	94\\
65	86\\
66	10\\
67	2\\
68	358\\
};
\addlegendentry{F+R}

\addplot [color=mycolor4, line width=2.0pt]
  table[row sep=crcr]{%
1	124\\
2	2681\\
3	27\\
4	27\\
5	23\\
6	9\\
7	3\\
8	3\\
9	3\\
10	3\\
11	46\\
12	3\\
13	3\\
14	11\\
15	3\\
16	9\\
17	373\\
18	272\\
19	61\\
20	115\\
21	109\\
22	58\\
23	35\\
24	34\\
25	23\\
26	113\\
27	115\\
28	103\\
29	47\\
30	35\\
31	31\\
32	83\\
33	111\\
34	101\\
35	63\\
36	33\\
37	37\\
38	39\\
39	103\\
40	6\\
41	21\\
42	61\\
43	37\\
44	39\\
45	37\\
46	33\\
47	35\\
48	37\\
49	35\\
50	37\\
51	37\\
52	37\\
53	35\\
54	37\\
55	35\\
56	37\\
57	19\\
58	33\\
59	37\\
60	37\\
61	35\\
62	37\\
63	35\\
64	33\\
65	33\\
66	27\\
67	3\\
68	33\\
69	372\\
};
\addlegendentry{P+P}

\end{axis}
\end{tikzpicture}%
\caption{Detail of a trace of misses measured each 100 $\mu$s for each of the approaches.} \label{rsa_time}
\end{figure}
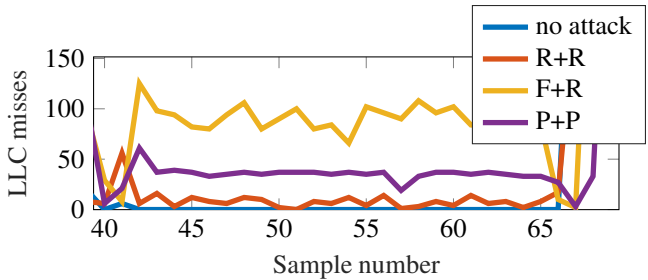

For this reason we have also monitored the victim LLC misses periodically, with a sampling rate of 100 $\mu$s. The results obtained in this case confirm that the main differences between RELOAD+REFRESH and the normal operation of the decryption process occur during the initialization steps. During this initialization, the number of misses caused by RELOAD+REFRESH and PRIME+PROBE is lower than in the normal execution. Later, when the number of misses of the normal operation tends to zero, the number of misses for the RELOAD+REFRESH is also close to zero. On the contrary, both FLUSH+RELOAD and PRIME+PROBE cause a noticeable amount of misses. Since detection mechanisms such as CacheShield \cite{Briongos_2018}, define a region in with some misses are tolerated to avoid false positives, our attack will not trigger an alarm. Figure \ref{rsa_time} shows the section of the decryption process in which the number of misses has become stable.

For each sample we get with the RELOAD+REFRESH approach, we have to perform both operations and then wait for 3000 cycles. The RSA decryption process runs in parallel, which means both victim and attacker can try to access the memory simultaneously. If the victim tries to execute the multiply operation when the attacker is flushing and reloading the mentioned line, the victim may get a miss. Therefore, a few misses can be observed in figure \ref{rsa_time} for our approach.

\section{Discussion of the results}

The absence of randomness in the replacement algorithm makes it possible to accurately determine which of the elements located in a cache set will be evicted in case of conflict. Also, the accurate timers included in Intel processors altogether with the \emph{cflush} instruction allow to trace accesses to the different caches and to force the cache lines to have the desired ages, we exploit these facts to run undetectable attacks.

RELOAD+REFRESH is just one way to exploit the eviction policy assuming some kind of memory sharing mechanism enabled. In the case that the victim and the attacker do not share memory, our proposal can be adapted so the attacker can track his own data. The attacker has to prepare the data in the set, in such a way that once the victim places his data in the cache, the eviction candidate is one of the elements the attacker controls. The target address will only be evicted if not used by the victim, who will see a cache miss the next time that tries to use it. If there is an encryption process going on, the most likely situation is that the victim uses the data. As a result, the number of misses will be limited. This is a new idea, that would lead to different results and a new attack, that we leave for future work.

The knowledge of the eviction policy, enables the usage of a different access pattern to gain the information about the victim and to ensure its data is really evicted from the cache, reducing the amount of false positives. Thus, PRIME+PROBE attacks, EVICT+RELOAD attacks or any attack requiring to evict some data from the cache can benefit from our results. Moreover, the PROBE step can, in some cases, be reduced to just one access to the eviction candidate. 

\subsection{Performance of Reload+Refresh}

As a difference with traditional cache attacks, one of our main objectives is to be stealthy. This involves we should cause as few cache misses and as low delay as possible on the victim process. Consequently, our sampling rate has to be at least the same as the rate at which the victim using its own data. We have already showed that it is possible to extract information referring to a RSA decryption, however in this section we discuss the performance of our approach and the resolution penalty paid as the price for stealthiness. Compared with traditional attacks such as PRIME+PROBE or FLUSH+RELOAD, our proposal is more sophisticated and involves extra steps in order to retrieve the same information. 

RELOAD+REFRESH includes an access to one address not located in the working set (the conflicting address) plus some others accesses focused in checking if the data has been replaced and in re-allocating the data in the desired places within the cache set. As a consequence, it induces some misses on the attacker side and the corresponding delays. In our test machine we have measured the time in cycles it takes to perform both the RELOAD (1012) and the REFRESH (447) operations as well as the mean time resulting of performing the two operations sequentially, that is, the maximum sampling rate (1520). When the number of ways per set increases, we do not expect variations in the RELOAD time, however there will be a proportional increase in the REFRESH time as it would be for a PRIME+PROBE attack.

Regarding to the capability of the channel to extract information, FLUSH+RELOAD attacks are the fastest and most accurate. RELOAD+REFRESH attacks do not lose accuracy, nevertheless, their performance is degraded due to the extra accesses. The mean times between samples that we have measured for the FLUSH+RELOAD attack is about 260 cycles, whereas this time is 810 cycles for PRIME+PROBE. Note that in the PRIME+PROBE approach the time varies between 280 and 1950, which is a great variation and makes it harder to accurately obtain samples with resolutions below 1950. 

To provide more insights into the resolution required to obtain data from algorithms such as RSA, we have gathered time measurements between calls to square and multiply. This time is approximately 1400 cycles for keys with 1024 bits and 3100 cycles for keys with 2048 bits. When the key length of RSA is 1024 bits, both PRIME+PROBE and RELOAD+REFRESH may have trouble getting the proper number of samples. However, we have been able to accurately retrieve 93\% of the multiply executions gathering the information at maximum speed with the RELOAD+REFRESH attack.

\section{Conclusion}
This work presented a thorough analysis of cache replacement policies implemented in Intel processors covering from 4th to 8th generations. To this end, we have developed a methodology that allows us to test the accuracy of each policy by comparing the data that such policy selects as the eviction candidate with the data truly evicted after forcing a miss.  

The RELOAD+REFRESH attack builds on this deep understanding of the platforms replacement policy to stealthily exploit cache accesses to extract information about a victim. We have demonstrated the feasibility of our approach by targeting AES and RSA and retrieving as much information as we can retrieve with other state of the art cache attacks. Additionally, we have have monitored the victim while running these attacks to confirm that our attack causes a negligible amount of last level cache misses, rendering it impossible to detect with current countermeasures.

These results are not only useful for broadening the understanding of modern CPU caches and their performance but also for improving previous attacks and eviction strategies. Our work also demonstrates that new detection countermeasures have to be designed to protect against RELOAD+REFRESH. 

\section*{Acknowledgment}
Visit of Samira Briongos to L\"ubeck has been supported by a grant from the Universidad Polit\'ecnica de Madrid.
This work was in part supported by the National Science Foundation under Grant No. CNS-1618837 and by the EU (FEDER), the Spanish Ministry of Economy and Competitiveness, under contract RTC-2016-6090-3, the Centre for the Development of Industrial Technology (CDTI), under contracts IDI-20171183 and IDI-20171194.

\bibliographystyle{plain}
\bibliography{ms}

\end{document}